\documentclass[useAMS,usenatbib]{mnras}
\usepackage{amsmath}
\usepackage{amssymb}
\usepackage{graphicx}
\usepackage{subfig}
\usepackage{float}
\usepackage{mathtools}
\newcommand\textlcsc[1]{\textsc{\MakeLowercase{#1}}}
\newcommand{\bff}{\bf}
\renewcommand{\bff}{}   
\def\jref@jnl#1{{\rm#1}}

\def\aj{\jref@jnl{AJ}}                   
\def\araa{\jref@jnl{ARA\&A}}             
\def\apj{\jref@jnl{ApJ}}                 
\def\apjl{\jref@jnl{ApJ}}                
\def\apjs{\jref@jnl{ApJS}}               
\def\ao{\jref@jnl{Appl.~Opt.}}           
\def\apss{\jref@jnl{Ap\&SS}}             
\def\aap{\jref@jnl{A\&A}}                
\def\aapr{\jref@jnl{A\&A~Rev.}}          
\def\aaps{\jref@jnl{A\&AS}}              
\def\azh{\jref@jnl{AZh}}                 
\def\baas{\jref@jnl{BAAS}}               
\def\jrasc{\jref@jnl{JRASC}}             
\def\memras{\jref@jnl{MmRAS}}            
\def\mnras{\jref@jnl{MNRAS}}             
\def\pasa{\jref@jnl{PASA}}
\def\pra{\jref@jnl{Phys.~Rev.~A}}        
\def\prb{\jref@jnl{Phys.~Rev.~B}}        
\def\prc{\jref@jnl{Phys.~Rev.~C}}        
\def\prd{\jref@jnl{Phys.~Rev.~D}}        
\def\pre{\jref@jnl{Phys.~Rev.~E}}        
\def\prl{\jref@jnl{Phys.~Rev.~Lett.}}    
\def\pasp{\jref@jnl{PASP}}               
\def\pasj{\jref@jnl{PASJ}}               
\def\qjras{\jref@jnl{QJRAS}}             
\def\skytel{\jref@jnl{S\&T}}             
\def\solphys{\jref@jnl{Sol.~Phys.}}      
\def\sovast{\jref@jnl{Soviet~Ast.}}      
\def\ssr{\jref@jnl{Space~Sci.~Rev.}}     
\def\zap{\jref@jnl{ZAp}}                 
\def\nat{\jref@jnl{Nature}}              
\def\iaucirc{\jref@jnl{IAU~Circ.}}       
\def\aplett{\jref@jnl{Astrophys.~Lett.}} 
\def\apspr{\jref@jnl{Astrophys.~Space~Phys.~Res.}}
\def\bain{\jref@jnl{Bull.~Astron.~Inst.~Netherlands}} 
\def\fcp{\jref@jnl{Fund.~Cosmic~Phys.}}  
\def\gca{\jref@jnl{Geochim.~Cosmochim.~Acta}}   
\def\grl{\jref@jnl{Geophys.~Res.~Lett.}} 
\def\jcp{\jref@jnl{J.~Chem.~Phys.}}      
\def\jgr{\jref@jnl{J.~Geophys.~Res.}}    
\def\jqsrt{\jref@jnl{J.~Quant.~Spec.~Radiat.~Transf.}}
\def\memsai{\jref@jnl{Mem.~Soc.~Astron.~Italiana}}
\def\nphysa{\jref@jnl{Nucl.~Phys.~A}}   
\def\physrep{\jref@jnl{Phys.~Rep.}}   
\def\physscr{\jref@jnl{Phys.~Scr}}   
\def\planss{\jref@jnl{Planet.~Space~Sci.}}   
\def\procspie{\jref@jnl{Proc.~SPIE}}   

\title[Disc heating mechanisms]{Vertical disc heating in Milky Way-sized galaxies in a cosmological context}
\author[Grand et al.]{\parbox[t]{\textwidth}{
Robert J. J. Grand$^{12}$\thanks{robert.grand@h-its.org}, Volker Springel$^{12}$, Facundo A. G\'{o}mez$^3$, Federico Marinacci$^4$, R\"{u}diger Pakmor$^1$, David J. R. Campbell$^5$ and Adrian Jenkins$^5$} \vspace{10pt} \\
$^1$Heidelberger Institut f\"{u}r Theoretische Studien, Schloss-Wolfsbrunnenweg 35, 69118 Heidelberg, Germany\\
$^2$Zentrum f\"{u}r Astronomie der Universit\"{a}t Heidelberg, Astronomisches Recheninstitut, M\"{o}nchhofstr. 12-14, 69120 Heidelberg, Germany\\
$^3$Max-Planck-Institut f\"{u}r Astrophysik, Karl-Schwarzschild-Str. 1, D-85748, Garching, Germany  \\
$^4$Department of Physics, Kavli Institute for Astrophysics and Space Research, MIT, Cambridge, MA 02139, USA \\
$^5$Institute for Computational Cosmology, Department of Physics, Durham University, South Road, Durham, DH1 3LE, UK\\
}
   
\date{Accepted XXX. Received YYY; in original form ZZZ}

\pubyear{2015}
    
\begin{document}

\label{firstpage}
\label{lastpage}
\pagerange{\pageref{firstpage}--\pageref{lastpage}}
\maketitle

\begin{abstract}
Vertically extended, high velocity dispersion stellar distributions appear to be a ubiquitous feature of disc galaxies, and both internal and external mechanisms have been proposed to be the major driver of their formation. However, it is unclear to what extent each mechanism can generate such a distribution, which is likely to depend on the assembly history of the galaxy. To this end, we perform 16 high resolution cosmological-zoom simulations of Milky Way-sized galaxies using the state-of-the-art cosmological magneto-hydrodynamical code \textlcsc{AREPO}, and analyse the evolution of the vertical kinematics of the stellar disc in connection with various heating mechanisms. We find that the bar is the dominant heating mechanism in most cases, whereas spiral arms, radial migration, and adiabatic heating from mid-plane density growth are all sub-dominant. The strongest source, though less prevalent than bars, originates from external perturbations from satellites/sub-halos of masses log$_{10} (M/\rm M_{\odot}) \gtrsim 10$. However, in many simulations the orbits of newborn star particles become cooler with time, such that they dominate the shape of the age-velocity dispersion relation and overall vertical disc structure unless a strong external perturbation takes place. 
\end{abstract}


\section{Introduction}

The existence of thickened stellar discs has been known since observations of the luminosity profiles of edge-on external galaxies \citep{B79,T79}  revealed an excess of starlight at more than $\sim1$ kpc vertical distance from the mid plane \citep{YD06,JIB08}.  Subsequently, \citet{GR83} showed that for our own Galaxy, the vertical density profile inferred from star counts in the solar neighbourhood is well fit by a superposition of two exponentials, leading to the popular interpretation that the stellar disc is made of two separate components of fundamentally different origin, although this scenario is contested  \citep[e.g.,][]{BRH12}. Whether stellar discs are composed of two distinct discs of different thickness, or one vertically extended distribution of stars, their origin and the principal mechanisms that govern their formation are not well understood. 

In the $\Lambda$CDM cosmogony, structure formation is predicted to proceed hierarchically through the clustering and merging of dark matter halos. In-falling baryonic matter condenses through radiative cooling at the centres of growing dark matter halo groups, which leads to the formation of massive galaxies surrounded by a distribution of smaller sub-halos  \citep{S77,WR78,MGG99}. In such a violent environment, there are many processes of external origin that are capable of generating a vertically extended distribution of stars in disc galaxies, for example, the accretion of tidally stripped stars from a satellite galaxy \citep{ANS03}, and kinematic heating of a pre-existing thin disc as a result of perturbations from satellite galaxies \citep{QHF93}. In addition, more violent mechanisms have been proposed, such as major mergers of gas rich systems \citep{Br04}. In this scenario, stars are born onto kinematically hot orbits at high redshift during starbursts associated with the merging system, generating a thickened disc of stars in a short time. A separate cosmologically motivated mechanism of rapid thick disc formation is that the bulk of stars associated with a thickened disc inherited their high velocity dispersions at birth from a turbulent, clumpy interstellar medium (ISM) during stages of heavy cosmological gas accretion at high redshift \citep{N98,BEM09}, which is later followed by more quiescent evolution after redshift $z=1$.  

Although major mergers have a significant effect on galactic structure, it is not clear whether such extreme events are required to create a thickened stellar disc. A key observable diagnostic for the formation history of the Galaxy is the age-velocity dispersion relation (AVR), whose shape is intimately related to past kinematic heating events and processes. However, the exact shape of the AVR remains a subject of debate, owing to the difficulties in measuring stellar ages \citep{So10}. {\bff For example, a feature that appears as a merger driven saturation in velocity dispersion, i.e., a step in the AVR \citep[e.g.,][]{QG01}, may be smoothed out to a power law after $30 \%$ age errors are considered \citep[e.g.,][]{MMF14}. Indeed, in} the solar neighbourhood, the Geneva-Copenhagen survey shows that the square of the vertical velocity dispersion increases linearly with age \citep{NMA04}\footnote{However, some authors argue that the relation may saturate at ages of $\sim 5$ Gyr \citep{DB98,AB09}, or that there is a jump in velocity dispersion $\sim 8-9$ Gyr ago \citep{SBM08}.}, which indicates that more gradual, secular heating processes play an important part in disc evolution. Among the sources of internal secular heating, several $N$-body simulation studies have indicated that spiral arms and bars are efficient at increasing the velocity dispersion of stars in the planar direction \citep{SC84,MQ06}, which can then be converted into vertical motion from star scattering off Giant molecular clouds (GMCs) \citep{C87}, other massive bodies, or in the form of vertical breathing waves induced by a bar and spiral arms \citep{MT97,FSF14,MFS15} or from satellite interactions \citep{GMO13}. Some evidence for a connection between bar strength and vertical heating has been found in $N$-body simulations of idealised isolated discs \citep{STT10} and in cosmological zoom simulations \citep{YS15}.

Aside from spiral-induced kinematic heating, \citet{LBK72} showed in analytical work that spiral arms are able to change the angular momentum value of individual stars \emph{without} increasing the random energy component of their orbits, in a process known as ``radial migration.'' \citet{SB02} studied radial migration in the context of idealised $N$-body simulations of stellar discs, and showed that star particles are radially redistributed over the whole disc as a result of many interactions with successive transient spiral arms. In the following years, numerical simulations proved to be an increasingly powerful tool for detailed study of the dynamical response of star particles to galactic structures. This led to the proposal of several mechanisms for radial migration, such as migration induced by spiral-bar coupling \citep{MF10}, wave mode coupling \citep{CQ12}, co-rotating spiral arms \citep{GKC11,GKC12,GKC13b,KHG14,HKG15} and migration by satellites\footnote{We note that stars that migrate via satellite interactions are kinematically heated, and predominantly sink toward the centre of the galaxy, which is different from radial migration from spiral arms.} \citep[][]{QMB09,BiKW12}. 

{\bff An important study of the impact of radial migration on the vertical disc structure is that of \citet{SB09b}, who examined the endpoint of a chemical evolution model with a parametrised form of radial migration. They concluded that the metal-rich, alpha-poor stars that originate from the inner regions of the Galaxy form a thick disc component around the solar neighbourhood, a result that hinges on the assumption that high velocity dispersion stars are equally as likely to radially migrate as kinematically cold stars, and in addition retain their high vertical energies in the disc outskirts.  However, in recent years it has been demonstrated in isolated simulations that a strong kinematic bias for migration towards kinematically cold stars exists \citep[e.g.,][]{VC14}, and that the vertical energy decreases for outward moving stars in accordance with the conservation of vertical action \citep{Min12,SoS12}. Recent $N$-body work has also speculated at the generation of thick discs via radial migration \citep[e.g.,][]{LRD11}, and it has been demonstrated that star particles that move from the inner to outer disc increase their vertical distance from the mid-plane, owing to the lower surface gravity and hence lower restoring force in the outer regions \citep{Rok12}. However, \citet{Min12} presented a detailed analysis of the radially migrated stars in $N$-body simulations of isolated discs and found that overall radial migration does not have a significant effect on galactic disc thickness, and may even cool the outer parts of the disc in the regime of strong satellite interaction \citep{MCM13,MCM14}. Nevertheless, the effects of radial migration on disc evolution remain heavily contested, and have been the subject of many studies \citep[e.g.,][]{H08,DiM13,MCM13,MCM14}.}

Nearly all numerical studies of detailed secular evolutionary processes to date simulate isolated disc systems with idealised initial conditions; simulations focussed solely on internal disc dynamics are set up with a perfectly symmetric disc in a closed box fashion, and those that include satellites are often set on contrived initial orbital parameters. Moreover, in order to achieve the numerical resolution required to suppress numerical heating \citep{Fu11} and resolve structures such as spiral arms, the dark matter halo component is often modelled by an analytical expression for the potential and not evolved self-consistently. These computational limitations prohibit the simulation of galaxy discs in a fully cosmological environment. However, full hydrodynamical cosmological simulations are now able to produce thin stellar discs with small bulges \citep{GBM10,ATM11,GC11,MBC12,AWN13,SB13,MPS14}, and are approaching the resolution required to obtain well-resolved sub-structure. 

In this paper, we present fully cosmological zoom simulations of Milky Way mass systems taken from a parent cosmological simulation at $z=127$ and evolved to $z=0$. The simulations include a wide range of galaxy formation physics, the employed models of which have been shown to reproduce realistic galaxy populations \citep[e.g.,][]{VGS13,MPS14}. We are therefore able to study complex dynamical phenomena in systems representative of a globally successful galaxy formation model, at the same level of detail as high resolution isolated, idealised simulations. 

This paper is organised as follows. In section~\ref{secsim} we describe the simulation technique and simulation suite. In section~\ref{secbs}, we quantify the spiral and bar strength evolution in all simulations and discuss their relation to disc heating. In section~\ref{secrm} we demonstrate that stellar radial migration is correlated to the strength of non-axisymmetric structure at different epochs throughout the evolution from $z=1$ to present day, and analyse the effects of radial migration on vertical disc structure. In section~\ref{secos} we discuss other sources of secular heating, such as adiabatic heating from inside-out disc growth and perturbations from satellites. We assess to what extent heating mechanisms drive the AVR in section~\ref{secav}. In section~\ref{secrs} we present a resolution study which shows that the disc heating in the simulations are not of numerical origin. Finally, we summarise our findings in section~\ref{seccon}.

\section{Methodology}
\label{secsim}

\subsection{Initial setup and simulation code}

In this section, we briefly describe the initial conditions and halo selection, which will be fully described in a forthcoming publication (Grand et al. in prep). The Milky Way-mass halo systems are selected from a parent dark matter only cosmological simulation. The simulation volume is a periodic cube of side length $100$ Mpc, with the standard $\Lambda$CDM cosmology. The adopted cosmological parameters are $\Omega _m = 0.307$, $\Omega _b = 0.048$, $\Omega _{\Lambda} = 0.693$ and a Hubble constant of $H_0 = 100 h$ km s$^{-1}$ Mpc$^{-1}$, where $h = 0.6777$, taken from \citet{PC13}. At the end-point of this simulation, candidate halos were selected within a narrow mass range interval around $10^{12} \rm M_{\odot}$ that are located at least $1.37$ Mpc from any object more than half the mass of the candidate, in order to select a sample of Milky Way size systems that are relatively isolated. 

We use the zoom-in technique to re-simulate the chosen systems at a resolution high enough to resolve galaxy formation and secular evolutionary processes. The initial conditions for the zoom simulations are generated by sampling the region in which the main galaxy forms with a high number of low mass particles, and surrounding regions with particles that grow progressively in mass with increasing distance from the main galaxy. Such particle sampling increases computational efficiency while the correct simulation of relevant external effects such as mass infall and the cosmological tidal field is maintained. Once the dark matter particle distribution initial conditions are set, gas is added by splitting each original dark matter particle into a dark matter particle and gas cell pair, and the masses assigned to each are determined from the cosmological baryon mass fraction. The dark matter particle and gas cell in each pair are separated by a distance equal to half the mean inter-particle spacing, and the centre of mass and centre of mass velocity is retained. 

\begin{table*}
\centering
\caption{Table of simulation parameters. The columns are 1) Model name; 2) Virial mass \protect\footnotemark; 3) Virial radius; 4) Stellar mass\protect\footnotemark; 5) Inferred decomposed stellar disc mass; 6) Radial scale length; 7) Inferred decomposed stellar bulge mass; 8) Bulge effective radius; 9) Sersic index of the bulge and 10) Disc to total mass ratio.}
\label{t1}
\begin{tabular}{c c c c c c c c c c}
\hline
Run & $M_{\rm vir}$ $(\rm 10^{12} M_{\odot})$ & $R_{\rm vir}$ (kpc) & $M_{*}$ $(\rm 10^{10} M_{\odot})$ & $M_{\rm d}$ $(\rm 10^{10} M_{\odot})$ & $R_{\rm d}$ (kpc) & $M_{\rm b}$ $(\rm 10^{10} M_{\odot})$ & $R_{\rm eff}$ (kpc) & $n$ & $D/T$  \\
\hline                                                                     
      Au 2-4M  &   1.91  &   261.76   &    7.05   &    4.63   &    5.84   &    1.45   &    1.34   &    0.99    &    0.76 \\
      Au 3-4M  &   1.46  &   239.02   &    7.75   &    6.29   &    7.50   &    2.10   &    1.51   &    1.06    &    0.75 \\
      Au 5-4M  &   1.19  &  223.09    &   6.72     &  4.38     &  3.80     &  1.95     &  0.87     &  1.02      &  0.69\\
      Au 6-4M  &   1.04  &  213.82    &   4.75     &  3.92     &  4.53     &  0.67     &  1.30     &  1.01      &  0.85\\
      Au 9-4M  &   1.05  &   214.22   &    6.10    &   3.57    &   3.05    &   2.03    &   0.94    &   0.84     &   0.64\\
     Au 12-4M &    1.09 &    217.12   &    6.01    &   4.33    &   4.03    &   1.48    &   1.05    &   1.07     &   0.75\\
     Au 15-4M &    1.22 &    225.40   &    3.93    &   3.14    &   4.00    &   0.39    &   0.90    &   1.02     &   0.89\\
     Au 16-4M &    1.50 &    241.48   &    5.41    &   4.77    &   7.84    &   1.00    &   1.56     &  1.18      &  0.83\\
     Au 17-4M &    1.03 &    212.77   &    7.61    &   2.67    &   2.82    &   4.14    &   1.11     &  0.71      &  0.39\\
     Au 18-4M &    1.22 &    225.29   &    8.04    &   5.17    &   3.03    &   1.98    &   1.06     &  0.79      &  0.72\\
     Au 19-4M &    1.21 &    224.57   &    5.32    &   3.88    &   4.31    &   1.02    &   1.02     &  1.13      &  0.79\\
     Au 21-4M &    1.45 &    238.64   &    7.72    &   5.86    &   4.93    &   1.48    &   1.36     &  1.29      &  0.80\\
     Au 23-4M &    1.58 &    245.27   &    9.02    &   6.17    &   4.03    &   2.42    &   1.26     &  0.94      &  0.73\\
     Au 24-4M &    1.49 &    240.86   &    6.55    &   3.68    &   5.40    &   2.18    &   0.93     &  0.9      &  0.63\\
     Au 25-4M &    1.22 &    225.30   &    3.14    &   2.59    &   6.30    &   0.76    &   2.44     &  1.69      &  0.77\\
     Au 27-4M &    1.75 &    253.81   &    9.61    &   7.21    &   4.21    &   1.70    &   0.92     &  1.00      &  0.81\\
\hline
\end{tabular}
\end{table*}

The typical mass of a high resolution dark matter particle is $\sim 3 \times 10^{5}$ $\rm M_{\odot}$, and the baryonic mass resolution is $\sim 4 \times 10^{4}$ $\rm M_{\odot}$. The comoving gravitational softening length for the star particles and high resolution dark matter particles  is set to $750$ pc, therefore the physical gravitational softening length grows with the scale factor, until to a maximum physical softening of $375$ pc is reached, at which time it is kept constant. The softening length of gas cells is scaled by the mean radius of the cell, with a minimum comoving softening of $750$ pc and maximum physical softening of $1.85$ kpc.

The zoom re-simulations are performed with the $N$-body, magneto-hydrodynamics code \textlcsc{AREPO} \citep{Sp10}, which we describe briefly here. For further detailed description, we refer the reader to \citet{Sp10}. \textlcsc{AREPO} is a moving-mesh code that follows the evolution of magneto-hydrodynamics and collision-less dynamics in a cosmological context. Gravitational forces are calculated by a standard TreePM method \citep[e.g.][]{SP05}, which itself employs a Fast Fourier Transform method for long range forces, and a hierarchical oct-tree algorithm \citep{BH86} for short range forces, together with adaptive time-stepping. 

To follow the magneto-hydrodynamics, \textlcsc{AREPO} utilises a dynamic unstructured mesh constructed form a Voronoi tessellation of a set of mesh-generating points (the so-called Voronoi mesh), that allows for a finite-volume discretisation of the Euler equations for magneto-hydrodynamics. The \textlcsc{MHD} equations are solved with a second order Runge-Kutta integration scheme with high accuracy least square spatial gradient estimators of primitive variables  \citep{PSB15}, which are improvements to the original version of \textlcsc{AREPO} \citep{Sp10}. 

A unique feature of \textlcsc{AREPO} is that the mesh can be transformed through a mesh re-construction at any time-step, which is not the case for standard grid-based methods. Each mesh construction ensures that each cell contains a given target mass (specified to some tolerance), such that regions of high density are resolved with more cells than regions of low density. Furthermore, the mesh generating points are able to move with the fluid flow velocity, such that each cell of the newly constructed mesh moves approximately with the fluid at each point. In this way, \textlcsc{AREPO} both overcomes the Galilean invariance problem and significantly reduces advection errors of large supersonic bulk flows, which occur in regular fixed mesh codes. The quasi-Lagrangian characteristic of the method makes it relatable to other Lagrangian methods such as Smoothed Particle Hydrodynamics (SPH), although several caveats of the SPH method are eliminated, for example, there is no artificial viscosity, and the hydrodynamics of under dense regions are treated with higher accuracy.

\subsection{Physics model}

We briefly summarise here some of the most important physical processes implemented in \textlcsc{AREPO}, and refer the interested reader to \citet{VGS13} and \citet{MPS14} for more details. 

Primordial and metal-line cooling with self-shielding corrections is enabled. Following the model of \citet{FG09}, a spatially uniform UV background field is included, which completes reionization at redshift $z\sim6$. The interstellar medium gas is modelled with an effective equation of state \citep{SH03}, in which star formation occurs stochastically provided the density of a gas cell is high enough, and the temperature is not higher than that inferred by the equation of state. Each star particle represents a single-age stellar population, the stellar mass distribution of which follows a Chabrier Initial Mass Function \citep{C03}. A stellar mass of $8\, \rm M_{\odot}$ is taken to define the mass threshold for supernova type II (SNII) and supernova type Ia (SNIa), and the $2\times$ higher normalisation of the time-delay distribution of SNIa from \citet{MPS14b} is adopted. At each time-step, the amount of mass and metals produced from SNII and SNIa per star particle is calculated, and is distributed into neighbouring gas cells. In addition to gas enrichment, stellar feedback in the form of winds is modelled. This is done by converting a probabilistically sampled star forming gas cell into a wind particle and launching it with a velocity that scales with the local dark matter velocity dispersion. The wind particle is briefly decoupled from the gas as it travels, interacting with other matter only through gravity until it passes into a gas cell below a density threshold, in which it re-couples and deposits the mass, momentum, metals and energy that it inherited from the gas cell from which it was launched. The amount of metals that the wind particle inherits from it's progenitor gas cell is controlled by the wind metal loading parameter, which describes how much of the gas cell metal content is passed to the wind particle and how much is given to surrounding gas cells. This parametrisation of wind outflows is required in order to reproduce both the stellar mass and oxygen abundances of low mass haloes \citep{PS13}.

\addtocounter{footnote}{-2}

\stepcounter{footnote}\footnotetext{Defined to be the mass inside a sphere in which the mean matter density is 200 times the critical density, $\rho _{\rm crit} = 3H^2(z)/(8 \pi G)$.} 

\stepcounter{footnote}\footnotetext{Defined as the stellar mass within $0.1$ times the virial radius.}

Magnetic fields are implemented following the method described in \citet{PakS13}. A homogeneous magnetic field is seeded at $z=127$, whose (comoving) strength is chosen to be $10^{-14}$ Gauss, and the direction chosen {\bff along the $Z$ direction}. The choice of direction and strength has been shown to have little effect on the evolution \citep{PMS14,MVM15}. The divergence cleaning scheme of \citet{PR99} is implemented to ensure that $\nabla\cdot\boldsymbol{\rm B}= 0$.

Black Holes are seeded, and grow through gas accretion and merger processes based on the model introduced in \citet{SMH05}. AGN feedback is implemented in the quasar-phase through thermal heating of the neighbourhood of the black hole. {\bff The model includes also a radio mode feedback channel, in which bubbles of hot gas gently heat the gaseous halo surrounding the central galaxy to compensate for its radiative losses in the X-ray band, following the relations presented in \citet{NF00}.}

\subsection{Simulations}
\label{sims}

\begin{figure*}
\includegraphics[scale=1.]{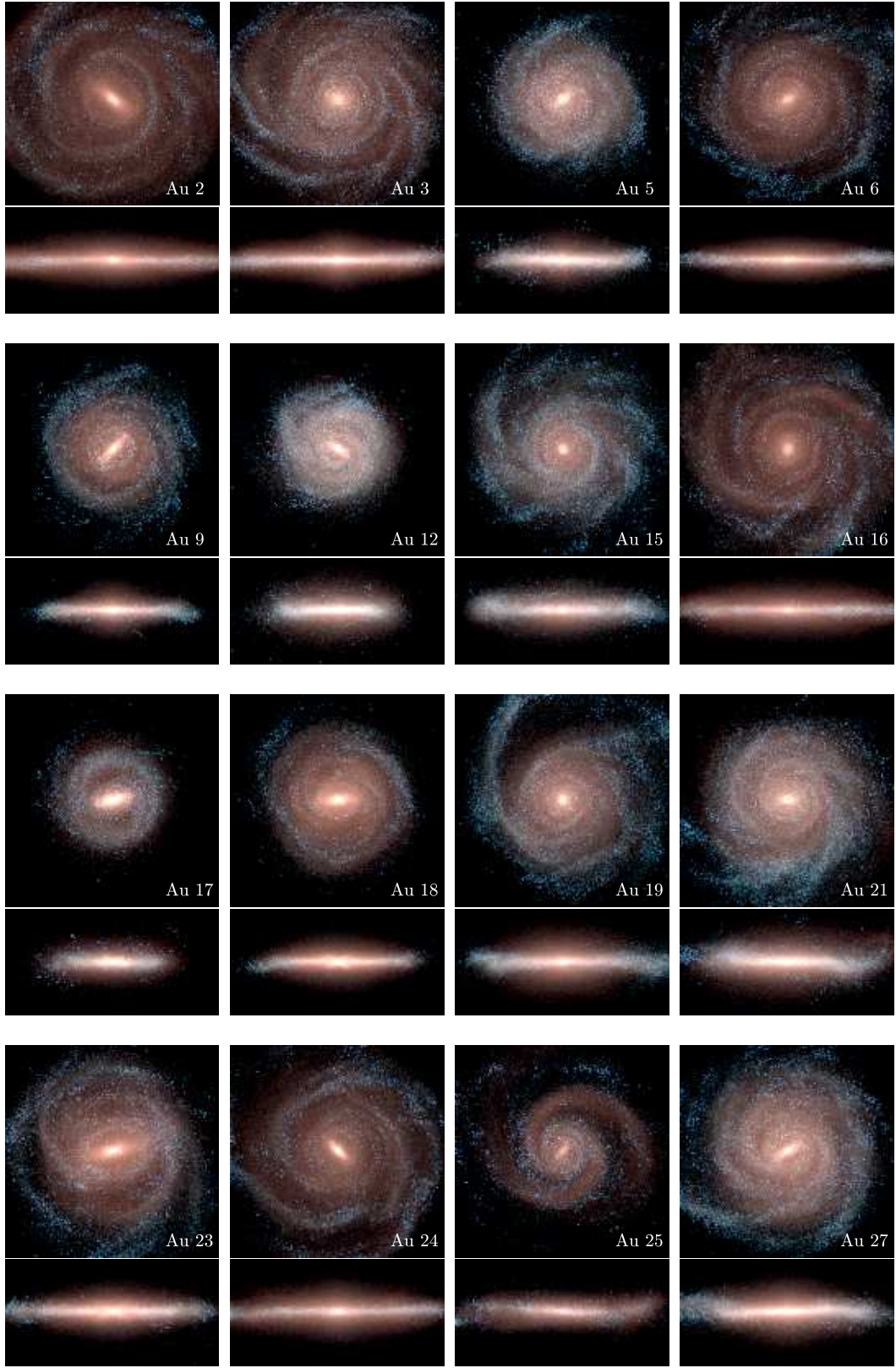}
\caption{The face-on and edge-on projected stellar density of each simulation at $z=0$. The images are synthesised from a projection of the $K$-, $B$- and $U$-band luminosity of stars, which are shown by the red, green and blue colour channels, in logarithmic intervals, respectively. Younger (older) star particles are therefore represented by bluer (redder) colours. The plot dimensions are $50\times 50 \times 25$ kpc.}
\label{proj}
\end{figure*}

\begin{figure*}\hspace{0.0mm}
\includegraphics[scale=1.1,trim={0 2.cm 0 0},clip]{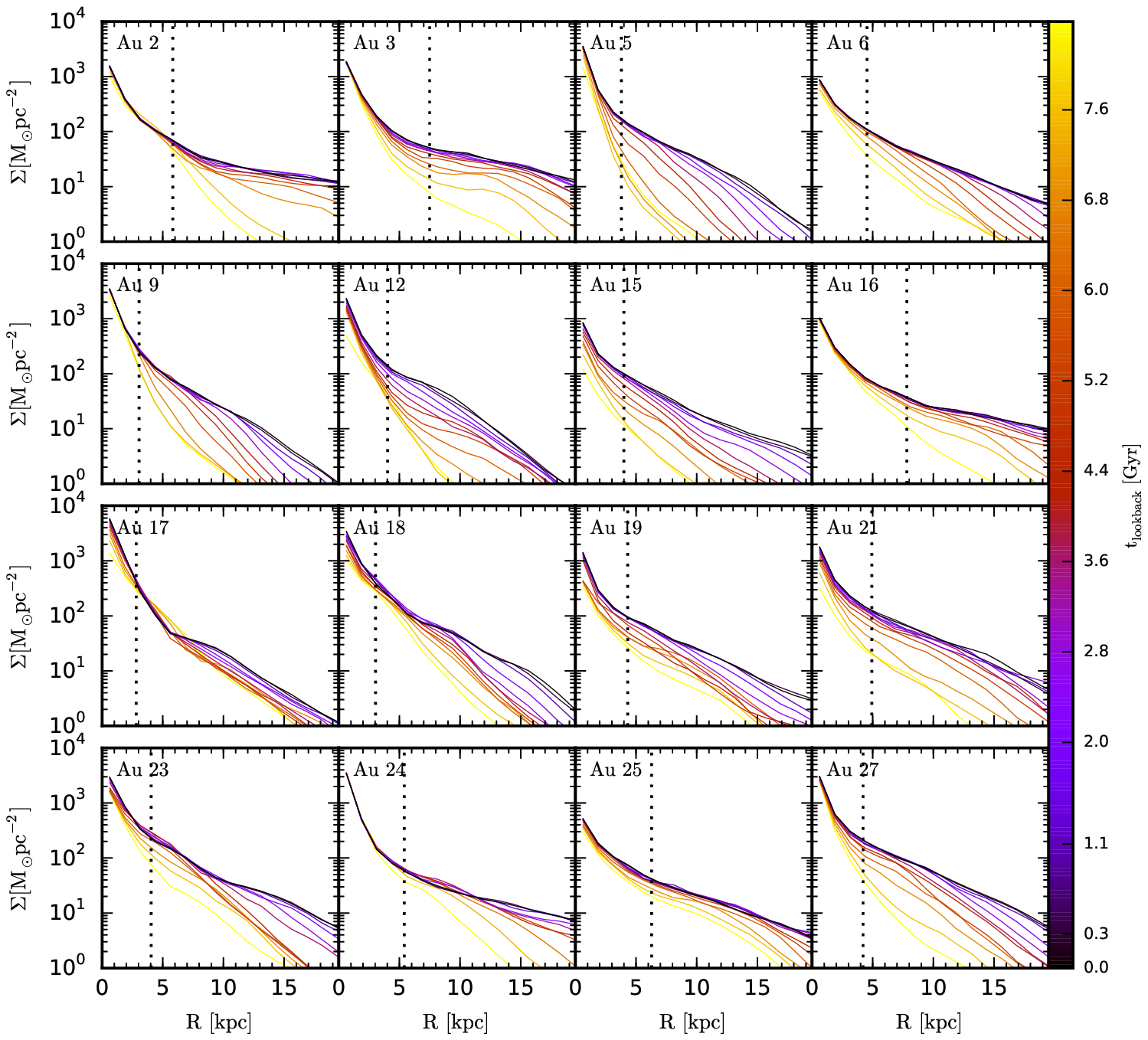}
\caption{The radial profile of the mass surface density of all star particles within $1$ kpc of the disc mid-plane, plotted at a series of times indicated by the colour bar. {\bff The disc scale length at present day is indicated by the dotted line.}}
\label{sden}
\end{figure*}

The current suite of simulations comprises 30 halos, the general properties of which will be fully analysed and compared to observations in a forthcoming publication (Grand et al. in prep). In this paper, we focus our analysis on a sub-set of these halos that exhibit clear stellar discs at present day, in order to elucidate vertical disc heating effects. The parameters of the resulting 16 halos are listed in Table 1. The halos take on the label ``Auriga'', which is abbreviated to ``Au''. The simulation names encode the halo number and resolution level, e.g., Au 2-4M indicates halo number 2 simulated at resolution level 4, which is consistent with the nomenclature of the Aquarius project \citep{SWV08}. The ``M'' indicates that magnetic fields are simulated. Because all simulations presented in this study are simulated at level 4 resolution with magnetic fields, for the remainder of this paper we abbreviate the simulation name to refer to only the halo number, e.g., Au 2.

In Fig.~\ref{proj} we show the present day face-on and edge-on projections of the stellar disc for all simulations, which have been re-orientated such that the disc plane is aligned with the $X$-$Y$ plane of our Cartesian coordinate system\footnote{To re-orient the disc plane, we compute and diagonalise the moment of inertia tensor of all star particles younger than 3 Gyr between {\bff 0} and 10 kpc radii. We tried several criteria for selecting star particles for computation of the inertia tensor, and found that this is a robust choice.}. The images are constructed by mapping the $K$-, $B$- and $U$-band luminosity of stars to red, green and blue colour channels, which indicates the distribution of old, intermediate and young stellar populations, respectively. In all cases, flat extended stellar discs and relatively small red spheroidal bulges are visible, the latter of which in most cases are so small as to barely protrude vertically out of the disc, for example, Au 16 and Au 19. Quantitatively, the surface density profile of the galaxies is decomposed into a disc and bulge component fit simultaneously with an exponential and Sersic profile  \citep[see][]{MPS14}. The inferred disc to total (baryonic) mass ratio ($D/T$) is listed in Table 1, which confirms the prominence of the stellar discs, and the present day radial scale length and bulge Sersic index reflect that the simulation suite contains a variety of discs similar in size and mass to that of the Milky Way \citep{RB13R,BR13}. Furthermore, the disc component in these halos shows evidence of ongoing star formation that tends to trace spiral/ring-shaped features, which is particularly clear in Au 3 and Au 19. In general, bar and spiral structure is well resolved, and clearly traces the underlying density field of the older populations of stars as well as sites of recent star formation. We therefore have a suitable simulation suite with which to study disc evolution in the context of galaxy disc heating processes.

\subsubsection{Disc surface density evolution}

We show the evolution of the surface density profile in Fig.~\ref{sden}, which clearly shows that the galaxy simulation sample includes a diverse set of formation histories. For example, several halos, such as Au 5, Au 9, and Au 15 grow steadily over a large radial range for the entire period of time shown, whereas Au 2 and Au 16 show very little increase in their surface density across the radial range examined {\bff after $t_{\rm lookback} \sim 6$ Gyr}. Some discs exhibit smooth and continual growth in the surface density around the disc edges, consistent with the concept that discs grow inside-out through continual acquisition of high angular momentum gas. Particularly clear examples of this type of growth include Au 3, Au 6 and Au 27, which show significant disc growth until times as late as $t_{\rm lookback} \sim 3$ Gyr. This is perhaps surprising, because galaxy discs are thought to be largely fully formed by $t_{\rm lookback} \sim 8$ Gyr ($z\sim1$) \citep{VDL13}. Analysis of the detailed formation history of these simulated galaxies will be presented in forthcoming publications. 

\begin{figure*}
\includegraphics[scale=1.1,trim={0 2.cm 0 0},clip]{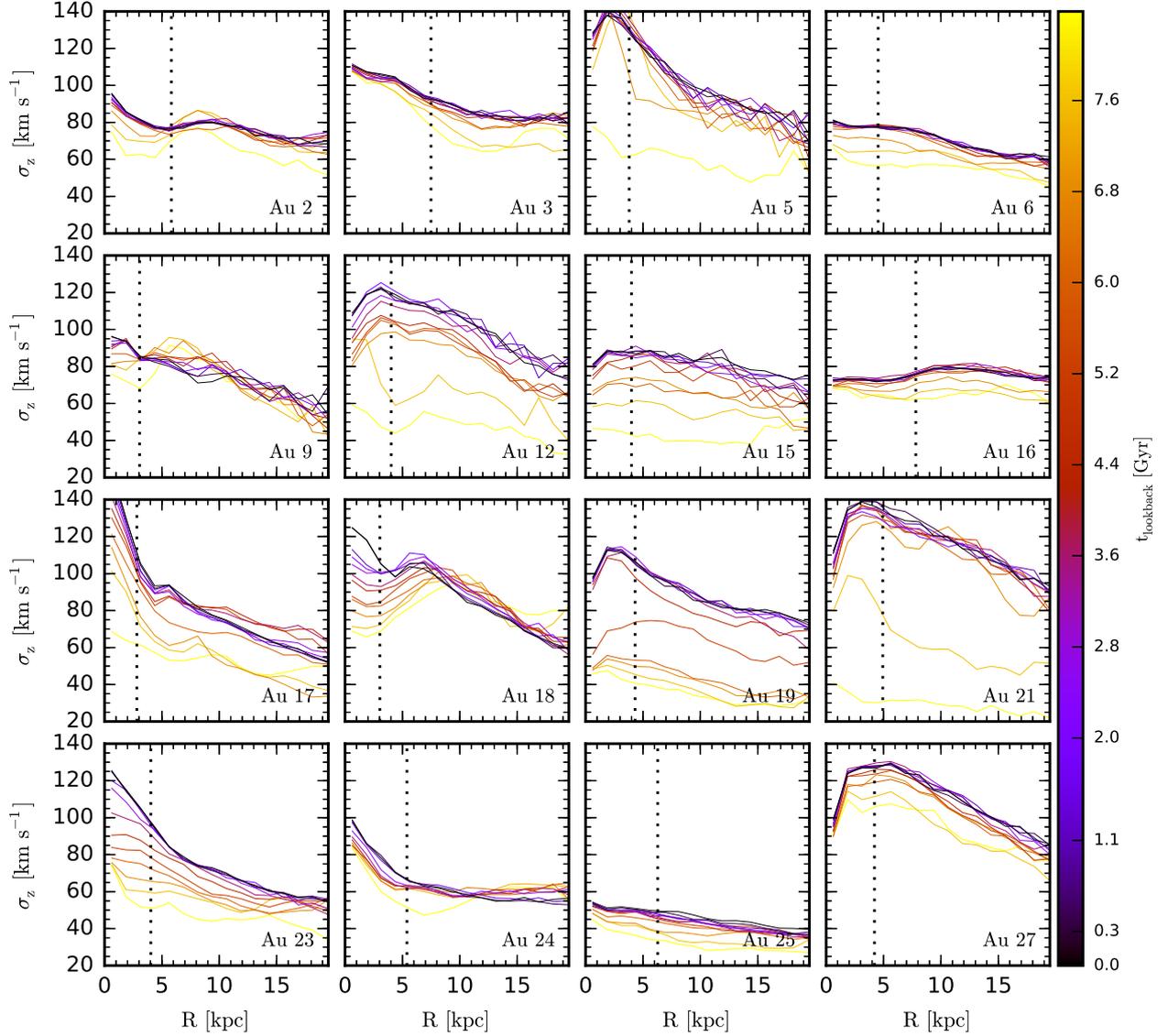}
\caption{The radial profile of the vertical velocity dispersion of a coeval population of star particles at a series of times indicated by the colour bar. For this coeval population, star particles are selected to be younger than $1$ Gyr at $t_{\rm lookback} = 8.5$ Gyr. {\bff The disc scale length at present day is indicated by the dotted line.}}
\label{monsigz}
\end{figure*}

\subsubsection{Vertical kinematic heating evolution}

{\bff To follow the kinematic heating rate of pre-existing star particles, we select, at  $t_{\rm lookback} = 8.5$ Gyr, a coeval population of star particles defined to be younger than $1$ Gyr. We calculate the radial profile of the vertical velocity dispersion of these star particles at a series of times between $t_{\rm lookback} = 8.5$ Gyr and} present day, which is shown in Fig.~\ref{monsigz} for all halos. Only star particles born in-situ are considered, in order to remove contamination of the kinematics by accreted stars. In nearly all simulations, the star particle population exhibits kinematic heating with time, however there is a variety of evolutions to be found within the sample. For example, {\bff Au 23 shows} a steady heating of the population between the radii $5$ and $20$ kpc, whereas Au 6 and Au 16 experience a relatively small amount of heating between a look back {\bff time of $8$ and $5$ Gyr} followed by very quiescent evolution. There are examples of strong, episodic heating in the evolution of Au 12 and Au 19. Interestingly, in some halos such as {\bff Au 9 and Au 18}, there is a radius inside of which the coeval population increases in vertical velocity dispersion, and outside of which the population appears to \emph{cool}. We will return to this feature of the velocity dispersion profile in section \ref{strm}. In what follows, we aim to identify the mechanism(s) responsible for the kinematic heating of pre-existing star particles highlighted in Fig.~\ref{monsigz}.

\section{Spiral and Bar structure}
\label{secbs}

\begin{figure*}
\centering \hspace{0.0mm}
\includegraphics[scale=1.15,trim={0 2.cm 0 0},clip]{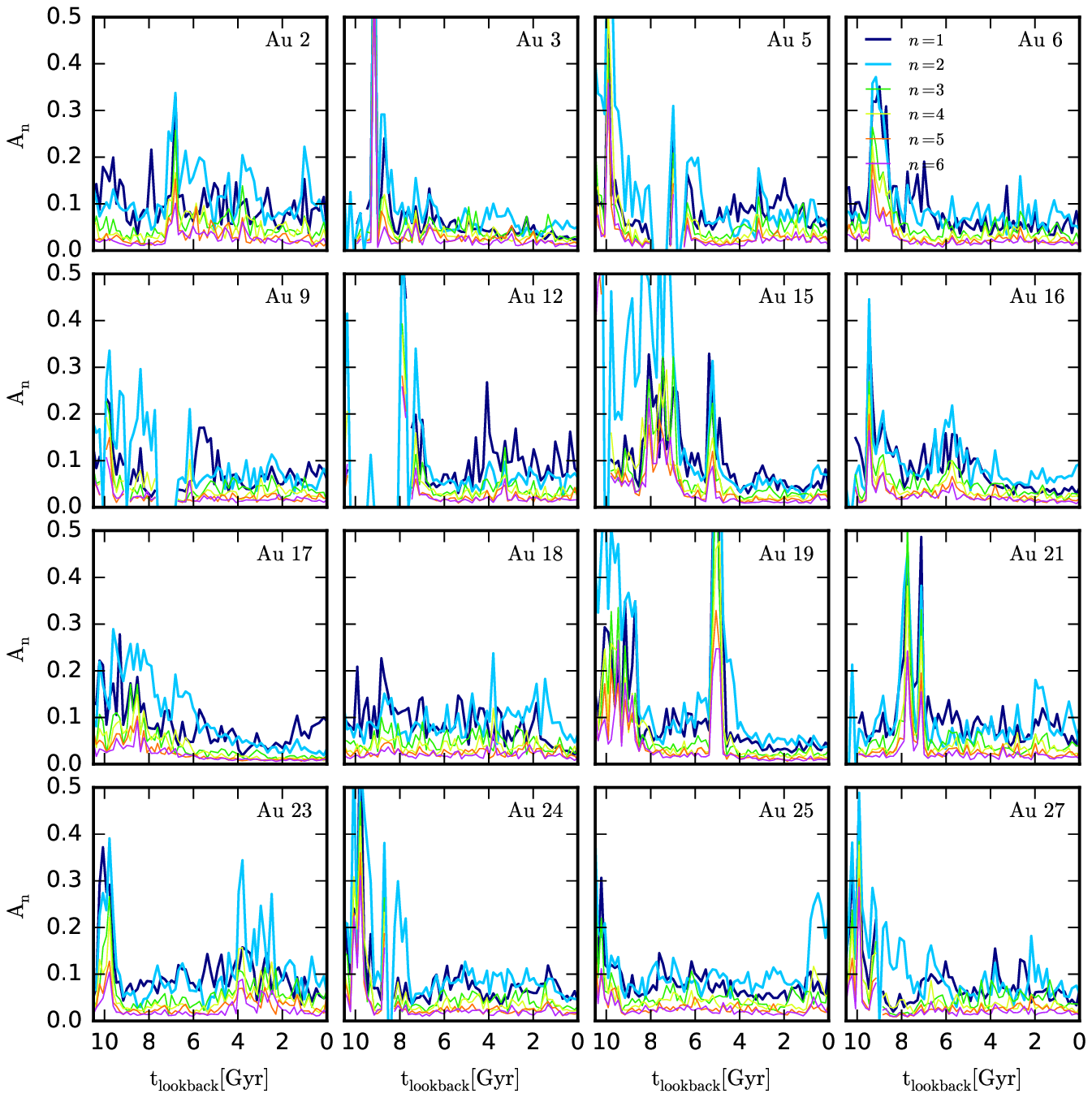}
\caption{The time evolution of the amplitudes of the Fourier modes $n=1$-$6$, calculated from equation (\ref{anmass}). The $n=1$ and 2 modes are represented by thick curves to emphasise their evolution.}
\label{spamp}
\end{figure*}

In order to establish a link between disc heating and bars and spiral arms, we must first quantify the strength of bar and spiral structure. We characterise the strength of spiral and bar structure by use of the Fourier mode approach \citep[e.g.,][]{QDBM10}. In a given radial annulus, the complex Fourier coefficients for an azimuthal pattern in the mass distribution with $n$-fold axi-symmetry are calculated as

\begin{equation}
C_n (R_j, t) = \sum _i ^N m_i e^{-in\theta _i},
\end{equation}
where $m_i$ and $\theta _i$ is the mass and azimuthal coordinate of the $i$-th star particle found within the $j$-th radial annulus in a set of 24 equally spaced radial annuli. The amplitude of the $n$-th Fourier mode in a given radial bin is then 

\begin{equation}
B_n (R_j,t) = \sqrt{a_n(R_j,t)^2 + b_n(R_j,t)^2},
\label{Bcf}
\end{equation}
where $a_n$ and $b_n$ are the real and imaginary parts of the complex Fourier coefficient, $C_n$. For a reliable individual quantification of the bar and spiral arms, it is important to determine at which radius the bar ends and spiral structure begins, such that appropriate radial ranges are chosen to calculate the Fourier amplitudes. We infer the semi-major axis length of the bar from consideration of the bar phase angle, given by

\begin{equation}
\theta ' _2 = \frac{1}{2} {\rm atan2} (b_2,a_2),
\label{pang}
\end{equation}
which returns a bar phase angle within the range $-\pi /2$ to $\pi / 2$ (owing to the $n=2$ rotational symmetry). Because the phase angle of the bar is generally constant over the bar radial range, we estimate the end of the bar semi-major axis as the radius at which the phase angle changes by more than $0.5$ rad \citep[similar to the criterion used by][]{GMM15}. The radial range considered for the calculation of the bar (spiral) strength is therefore taken to be $R < R_{\rm bar}$ ($R_{\rm bar} < R < 20$ kpc). We impose a lower limit for $R_{\rm bar}$ to be $R_s$, in order to prevent poor sampling in very small radial ranges.

The mass-weighted mean of the amplitudes of a given Fourier mode in a given radial range is

\begin{equation}
A_n (t) = \frac{\sum _j   B_{n} (R_j,t) }{\sum _j  B_{0} (R_j,t)},
\label{anmass}
\end{equation}
{\bff where $B_0$ is given by equation (\ref{Bcf}) for $n=0$.} Fig.~\ref{spamp} shows the spiral amplitude, $A_{n}$, for $n=1$-$6$ as a function of time for all halos. In most halos, the most dominant Fourier mode is the $n=2$ mode, which corresponds to a two-armed spiral structure. This is corroborated from visual inspection of the face-on stellar projections in Fig.~\ref{proj}, which illustrates that most of the discs show a grand design type of spiral morphology, as opposed to the many armed flocculent type of spiral structure. Fig.~\ref{spamp} shows that $n=1$, lop-sided modes are prominent features also, which in some cases, such as Au 5 and 12, dominate over even the $n=2$ modes for much of the evolution. The amplitudes of the dominant Fourier modes oscillate with time around amplitudes of $\sim 0.05$-$0.1$, which is similar to what is found in idealised simulations of isolated discs in which transient, recurring spiral arms continually form and disrupt \citep[e.g.,][]{GKC11,WBS11,BSW12,GKC13,S14,DB14}. It is evident also that there are specific times at which the amplitudes of all Fourier modes rapidly spike and decay, which appear to be a result of external perturbations from satellites that excite spiral structure. 

\begin{figure*}
\centering
\subfloat{\includegraphics[scale=2.,trim={0 0.5cm 0 0.5cm},clip]{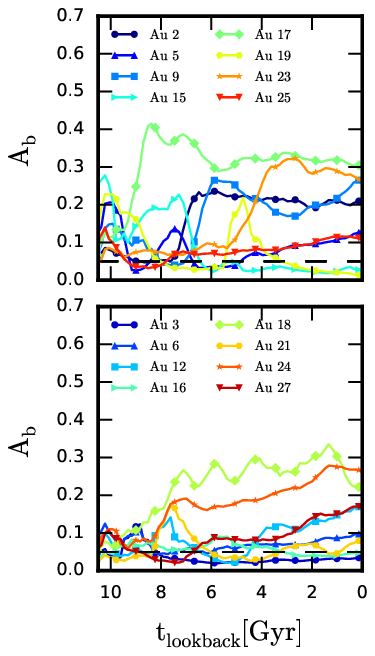}}
\subfloat{\includegraphics[scale=2.,trim={0 0.5cm 0 0.5cm},clip]{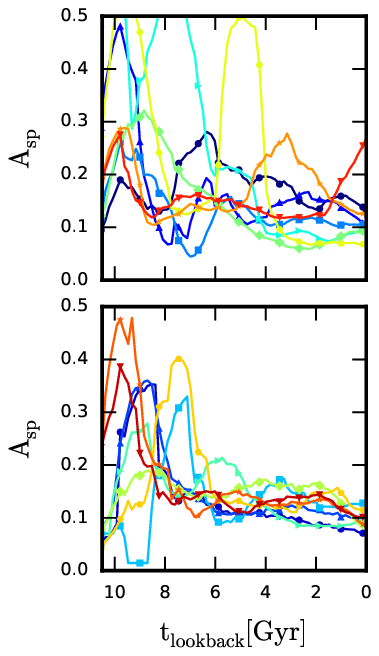}}
\caption{\emph{Left panels}: The bar strength, characterised by amplitude of the $n=2$ Fourier mode calculated from equation (\ref{anmass}) within the bar radius given by equation (\ref{pang}), as a function of time for all the simulations. A moving average of $0.5$ Gyr width is applied to smooth out rapid fluctuations. The amplitude value of 0.05 is marked by the dashed line. \emph{Right panels}: The amplitude of the spiral structure, characterised by the sum in quadrature of the $n=1$-6 Fourier mode amplitudes calculated from equation (\ref{anmass}) in the radial range between the bar radius and 20 kpc, as a function of time for all simulations. A moving average of $1.0$ Gyr width is applied to smooth out rapid fluctuations. The simulation sample is spread over two sets of panels (top and bottom row), in order to aid visual inspection.}
\label{sbev}
\end{figure*}


\begin{figure*}
\centering
\subfloat{\includegraphics[scale=0.9,trim={0 1.5cm 0 1.2cm},clip]{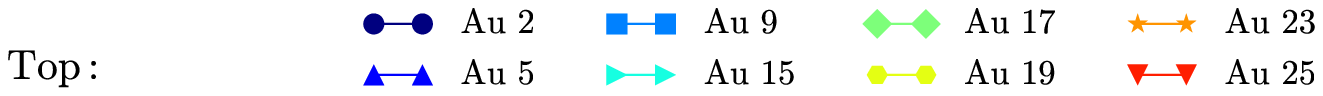}}\\ 
\subfloat{\includegraphics[scale=0.9,trim={0 1.5cm 0 1.2cm},clip]{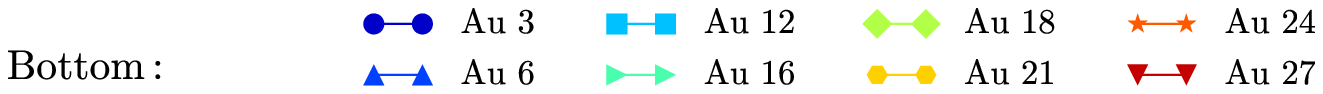}}\\ 
\subfloat{\includegraphics[scale=1.7,trim={0 0.5cm 0 0.75cm},clip]{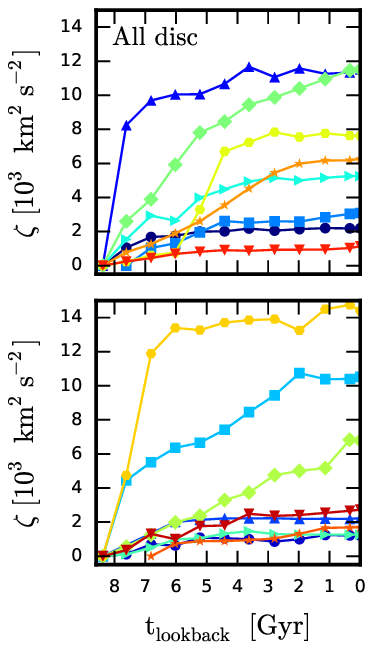}} 
\subfloat{\includegraphics[scale=1.7,trim={0.9cm 0.5cm 0 0.7cm},clip]{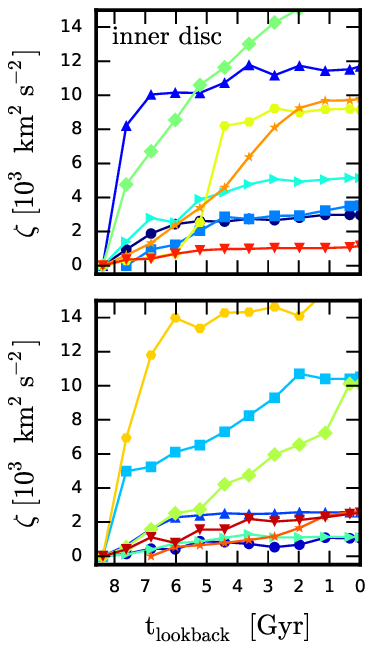}}
\subfloat{\includegraphics[scale=1.7,trim={0.9cm 0.5cm 0 0.7cm},clip]{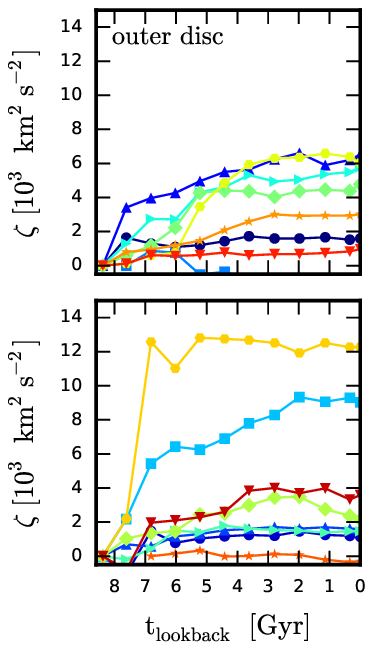}}\\
\caption{\emph{Left panels}: {\bff The change in vertical kinetic energy ($\zeta$) of the coeval star particles found in the disc born at a lookback time between $8.5$ and $9.5$ Gyr, plotted as a function time. \emph{Middle panels}: The same as the left panels but for star particles in the inner disc ($R < R_{\rm d}$). \emph{Right panels}: The same as the left panels but for star particles in the outer disc ($2R_{\rm d} < R < 3R_{\rm d}$). The simulation sample is spread over two sets of panels (top and bottom row), in order to aid visual inspection.}}
\label{sbheat}
\end{figure*}

\begin{figure*}
\centering
\subfloat{\includegraphics[scale=0.9,trim={0 1.5cm 0 1.2cm},clip]{figures/sim_legend_wide_top}}\\ 
\subfloat{\includegraphics[scale=0.9,trim={0 1.5cm 0 1.2cm},clip]{figures/sim_legend_wide_bot}}\\ 
\subfloat{\includegraphics[scale=1.7,trim={0 0.5cm 0 0.7cm},clip]{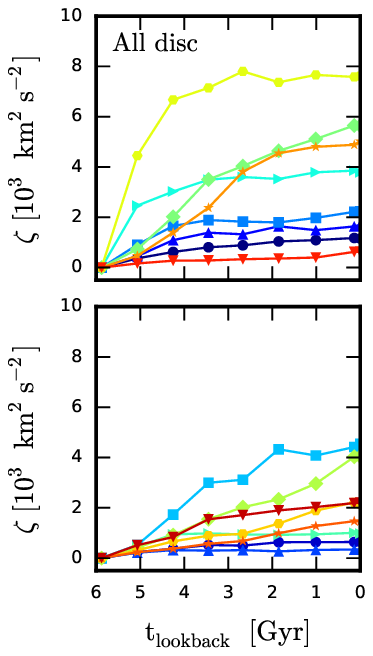}} 
\subfloat{\includegraphics[scale=1.7,trim={0.9cm 0.5cm 0 0.7cm},clip]{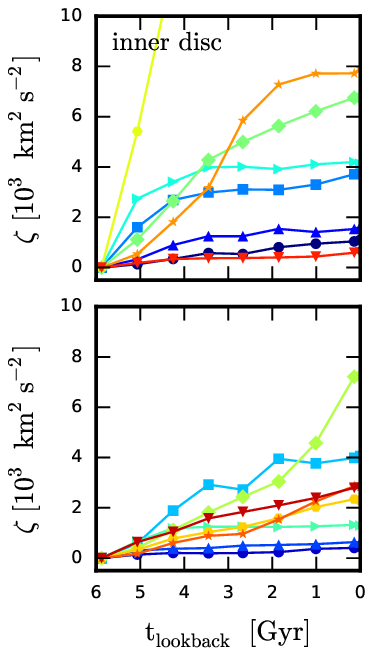}}
\subfloat{\includegraphics[scale=1.7,trim={0.9cm 0.5cm 0 0.7cm},clip]{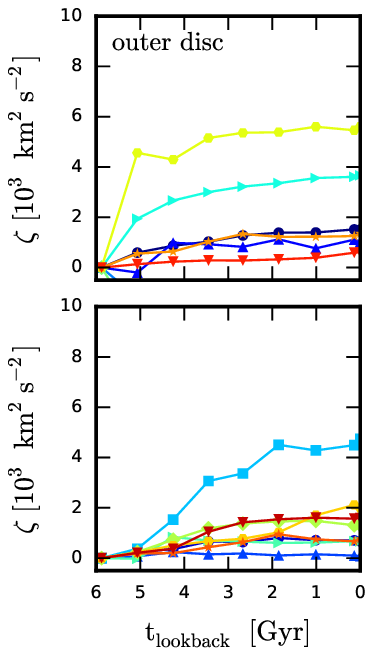}}
\caption{As Fig.~\ref{sbheat} but for a coeval population selected between $t_{\rm lookback} = 6$ and 7 Gyr.}
\label{sbheat2}
\end{figure*}


The left panels of Fig.~\ref{sbev} show the evolution of bar strength for all the simulations, with a moving average of width 0.5 Gyr applied in order to erase rapid fluctuations. We see that if we consider a present day amplitude of less than 0.05 to be the threshold to determine the presence of a bar in a given disc, our sample of 16 halos contains 12 barred-spiral discs and 4 spiral-only discs. Furthermore, it is clear that there is a wide variety of bar strength evolution in the barred discs: there is a spread in the time at which bars begin to grow above the amplitude threshold, for example the bar in Au 5 and Au 6 begin to grow at a lookback time of $\sim5$ Gyr, whereas the bars in Au 2 and Au 17 exhibit an amplitude greater than $0.1$ at that time. There is a general tendency for bars that begin to grow at early times to evolve to the strongest bars at present day, however this is complicated by the various evolutionary histories. Most notably, there are three types of evolution: a) bars that show continual growth such as Au 12 and Au 27; b) bars that appear to remain at a given strength for long periods of time such as Au 2; c) and bars that exhibit a period of growth followed by a decline, followed by a resumption of growth such as Au 9.

\subsection{Effect of bars and spiral arms on disc heating}

To aid comparison of the vertical kinematic evolution with the evolution of non-axisymmetric structure, we calculate the total kinetic vertical energy of the same coeval star particles as shown in Fig.~\ref{monsigz} as a function of time, 

\begin{equation}
\zeta = \sigma _z ^2 (t) - \sigma _z ^2 (t_0),
\label{zeta}
\end{equation}
{\bff where $t_0 = 8.5$ Gyr} lookback time. The left panels of Fig.~\ref{sbheat} show for all simulations the evolution of this quantity {\bff calculated from all the star particles within $0.1 R_{\rm vir}$, which provides a measure of the total kinematic heating over the disc. Several cases of high degrees of heating are evident, which can be categorised as sudden, e.g., Au 5, Au 19 and Au 21, and smooth and continuous, e.g., Au 17, Au 18 and Au 23. Some halos exhibit remarkably low kinematic heating over this period, for example, Au 25.}

By comparing the bar strength evolution (left panels of Fig.~\ref{sbev}) with the vertical kinetic energy evolution (left panels of Fig.~\ref{sbheat}), we see that the most vigorously heated discs, {\bff Au 5}, Au 19 and 21, do not host particularly strong bars during the time period examined, nor during times {\bff prior to} strong heating, e.g., $t_{\rm lookback} \sim 5$ Gyr for Au 19. With the exception of these halos, however, there is evidence that bar strength is correlated with disc heating: the discs in {\bff Au 17, Au 18 and Au 23} possess the strongest bars in the simulation sample from $z=0.5$ ($t_{\rm lookback} = 5.2$ Gyr) to present day, and show the largest amount of heating of their respective coeval populations. Particularly interesting is the sudden heating of the disc in Au 18 at $t_{\rm lookback} \sim 1$ Gyr, which is accompanied by the equally sudden drop in bar strength at the same time. This corresponds to a bar buckling event that produces an X-shaped bulge (visible in Fig.~\ref{proj}). Such events are reported to cause stars originally on bar supporting orbits to transition suddenly to box-orbits that extend vertically out of the plane, and thus contribute to an increase in vertical energy \citep[e.g.,][]{P84,A12}. 

{\bff To further explore the link between bars and heating, we follow the evolution of $\zeta$ in the inner and outer disc, defined as $R<R_{\rm d}$, and $2R_{\rm d} < R < 3R_{\rm d}$ respectively, shown in the middle and right columns of Fig.~\ref{sbheat}. The heating of the halos with the strongest bars (Au 17, 18 and 23) is increased in the inner regions, and low in the outer regions, in line with the expectation that bars are most effectively heating the inner disc material. The heating rate of the halos that host the next three strongest bars, Au 2, 9 and 24, is also increased (decreased) in the inner (outer) region, though noticeably lower than the three strongest bar cases. As noted by several authors \citep{SeG07,AB09}, stars older than $\sim5$ Gyr are less sensitive probes of heating mechanisms than are younger stars, owing to possible saturation of their velocity dispersion. Therefore, we follow also the evolution of a younger coeval population shown in Fig.~\ref{sbheat2}, selected at $t_{\rm lookback} = 6$ Gyr, which is more responsive to late time evolution when bars are most prominent. In particular, the steep and steady increase that occurs for Au 24 in the inner regions after $t_{\rm lookback}\sim 4$ Gyr is consistent with the  growth of the bar to amplitudes of $>0.2$.}

Although these trends indicate that the bar is a prominent source of disc heating, it is clear that another source is required in order to account for the heating of discs with a weak/no bar, in particular halos Au 5, Au 15, Au 19 and Au 21, which exhibit {\bff stages of} significant heating that {\bff seem to be driven by a different/additional mechanism.} 

Another possible cause of disc heating is spiral structure, which has been shown to cause an increase in the radial and tangential velocity dispersion of stars. The increase in planar random motion is thought to be able to be converted into vertical motions by scattering from Giant Molecular Clouds \citep[GMCs, e.g.][]{C87} and/or breathing modes induced by non-axisymmetric structure \citep{FSF14,D14}. To aid comparison of the vertical kinetic energy evolution with the evolution of spiral structure strength, we show in the right panels of Fig.~\ref{sbev} the spiral amplitude of the $n=1$-$6$ Fourier modes summed in quadrature and smoothed with a moving average to enhance the clarity of the evolution. {\bff There is some evidence to suggest that spiral arms are linked to kinematic heating in the outer disc regions. For example, at times later than $t_{\rm lookback} \sim 6$ Gyr, halos Au 2 and 18, which exhibit relatively high spiral amplitudes of more than $0.15$, show some moderate heating in the outer disc (see right panels of Fig.~\ref{sbheat2}), whereas the low spiral amplitudes halos Au 3, Au 6 and Au 17 show very little heating (and indeed cooling in the case of Au 17). However, the disc} in Au 25 stands out as having the most clearly defined two-armed spiral structure (see Fig.~\ref{proj}) at late times, but the disc heating evolution is among the shallowest in the sample {\bff for all regions of the disc.} 

{\bff A complication in attributing spiral structure to heating in the outer disc is that many of the prominent spiral signatures seen in Fig.~\ref{sbev} seem to be caused by} satellite interactions (see Section \ref{extp}), which are likely to be the source of heating in these cases. {\bff A clear example of this is in the weakly-barred disc of Au 21, in which a sudden increase in spiral amplitude at $t_{\rm lookback} \sim 2$ Gyr coincides with an increase in $\zeta$. Aside from these difficulties in determining a correlation, the generally low level of heating that can be associated to spiral arms leads us to conclude that spiral structure is not a} dominant source of kinematic heating. That the spiral structure seems unlikely to be a source of disc heating in these simulations is not surprising, because the adopted ISM model does not currently resolve a cold, clumpy medium, therefore the conversion of radial motions to vertical motions by molecular cloud scattering is not possible.

\section{Radial migration}
\label{secrm}

\begin{figure*} \hspace{-10.0mm}
\includegraphics[scale=1.15,trim={0 2.cm 0 0},clip]{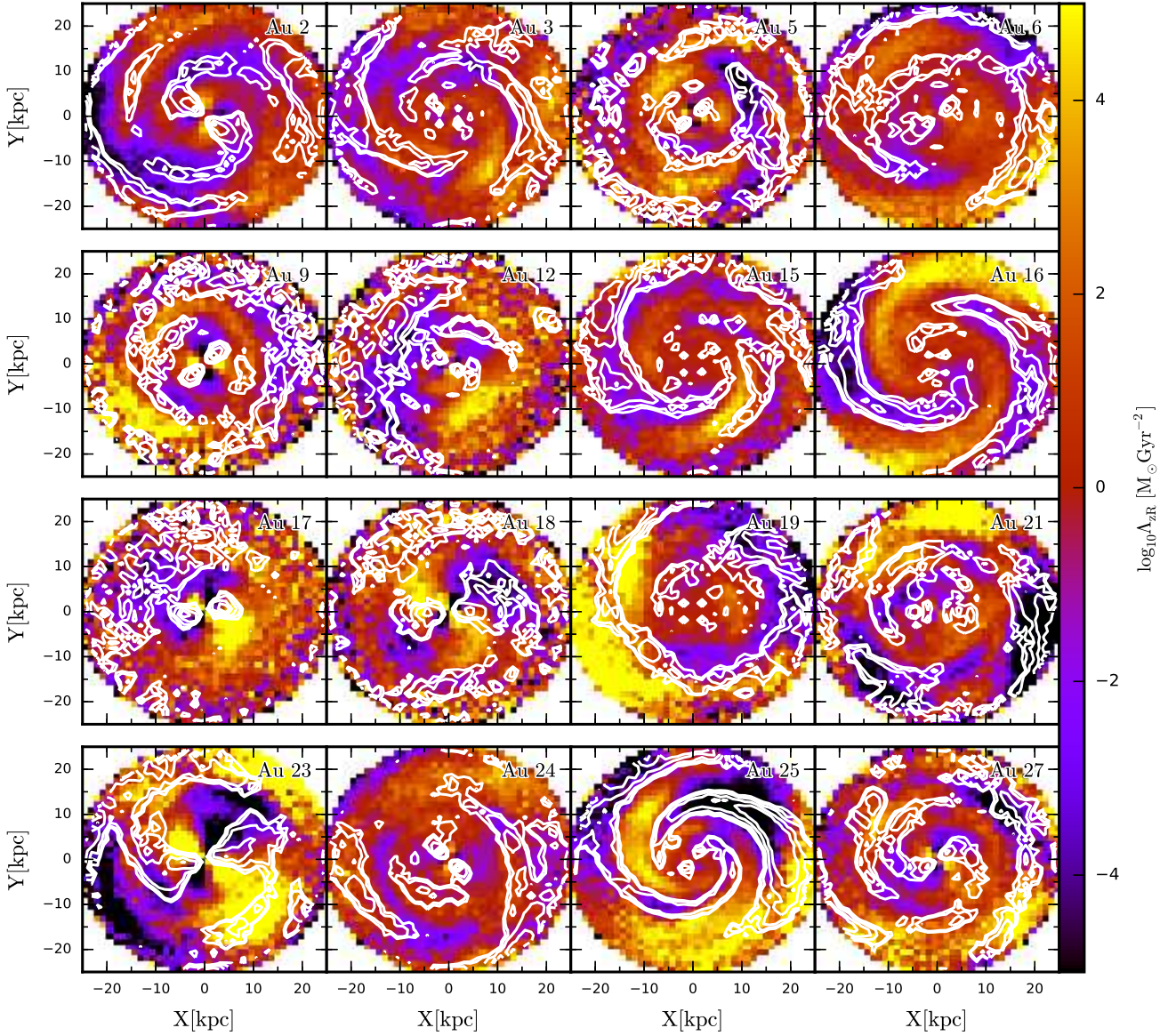}
\caption{The angular momentum flux tensor component, $\Lambda _{zR}$, as expressed in equation (\ref{amflux}), plotted on a 2D plane in configuration space aligned with the $X$-$Y$ plane, for all halos. Positive values indicate angular momentum flux in the outward radial direction, and negative values in the inward radial direction. White contours indicate the 0.05, 0.1, 0.2 and 0.3 levels of stellar azimuthal over-density.}
\label{lambda}
\end{figure*}

In this section, we investigate the effects of radial migration from bar and spiral structure on the vertical structure of the disc, which has been recently debated \citep[e.g.][]{SB09b,Min12,MMF14}. For this analysis, we focus on star particles that belong to the disc, which we select according to the orbital circularity. Following the method presented in \citet{MPS14}, we calculate the $z$-component of angular momentum, $L_z$, for each star particle, and define the circularity parameter as 

\begin{equation}
\epsilon = \frac{L_z(E)}{L_{z,\rm max}(E)},
\end{equation}
where $L_{z,\rm max}(E)$ is the maximum angular momentum allowed for the binding energy, $E$, of the star particle. Those that satisfy $\epsilon > 0.7$ are classified as disc star particles. This selection criterion ensures that we remove contamination from halo and bulge stars, and that we are left with a component of star particles with a well defined angular momentum along the $z$-axis. 

\subsection{Quantifying radial migration}

Before we investigate the impact of radial migration on the vertical disc structure, we first establish that angular momentum changes of individual star particles are attributed to bar and spiral structure. This requires a quantification of the angular momentum changes/strength of radial migration. We compute two separate quantities to characterise angular momentum changes of star particles: i) the angular momentum flux tensor, ii) second order diffusion coefficients in angular momentum space. The former has the advantage that it can be calculated at an instant in time on a grid in configuration space to give an indication of the directional flows of angular momentum. The latter encapsulates the radial migration that has occurred in a given time interval, and therefore reliably quantifies the amount of radial migration that has taken place over a period of galactic evolution.

\subsubsection{Angular momentum flux tensor}

The angular momentum flux tensor describes the current of a given component of angular momentum through a surface whose norm is perpendicular to the direction of the angular momentum component, and has the form \citep{BT08}

\begin{equation}
\Lambda _{im} = \epsilon_{ijk} x_j  \rho \overline{v_kv_m},
\end{equation}
where $\rho$ is the mass volume density, $\epsilon _{ijk}$ is the permutation tensor of an orthogonal basis set and $x$ and $v$ are position and velocity in the direction indicated by the sub-script. For our purposes, the angular momentum flux tensor is a useful quantity to calculate because it provides information of the angular momentum flow at an instant in time, which then may be correlated to galaxy structure at the same time and position in configuration space. For radial migration, the $z$-component of angular momentum that flows through a surface in the radial direction is the relevant element of the flux tensor, which is given as

\begin{equation}
\Lambda _{zR} = R  \rho \overline{v_{\phi}v_R} ,
\label{amflux}
\end{equation}
and is readily computable from a simulation snapshot. Fig.~\ref{lambda} shows this quantity calculated in bins of side length $1.25$ kpc, in the $X$-$Y$ plane for each of the simulations at $z=0$. It is clear that in some halos the direction of angular momentum flux traces out spiral structure in Fig.~\ref{proj}, for example, the flux tensor map of Au 16 shows a clear two-armed spiral shape that correlates well with the two-armed spiral over-density. In addition to spiral arms, bar-driven radial migration is evidenced by the clear quadrupole pattern near the centre of the barred galaxies, particularly in Au 23, which possesses one of the strongest bars in the suite at $z=0$. The azimuthally dependent angular momentum flux patterns are consistent with the systematic radial motion of star particles as a response to continual tangential gravitational forces supplied by transient spiral arms seen in isolated simulations \citep{GKC11,GKC12,GKC13b}, and indicative that spiral arms and a bar drive radial migration in our suite of simulations.

\subsubsection{Radial migration as angular momentum diffusion}

In a statistical context in which an ensemble of star particles is considered, radial migration can be approximated as a diffusion process \citep[e.g.,][]{BCP11,KPA15,SB15}. Over time, the distribution of a group of star particles located initially in a narrow range of angular momentum, $L_z$ (or guiding centre, $R_g$), will become broader as tangential forces act to change the guiding centre radii of the star particles. If the distribution in guiding centre (angular momentum) space can be approximated as a Gaussian, then the rate of change of the dispersion of the distribution can be taken to obey the spatial diffusion equation in cylindrical coordinates, the general solution of which is

\begin{equation}\label{c}
C(R_g, t) = \int _0 ^{\infty} A(k) e^{-D k^2 t} J_0(kR_g) k \mathrm{d} k ,
\end{equation}
where $J_0$ is the Bessel function of the first kind, $C$ is the distribution of star particles in guiding centre radius at a time, $t$, $k$ is a length scale and $D$ is the second order diffusion coefficient, which we assume to be time-independent for a given ensemble of star particles. From the assumption of the initial narrow star particle distribution in $R_g$, i.e., $C(R_g,0) = C_0 R_{g0} \delta (R_g - R_{g0})$, the Hankel transform of $C$ gives

\begin{equation}
\begin{split}\label{a}
A(R) & = C_0 R_{g0} \int _0 ^{\infty}  \delta (R_g - R_{g0}) J_0 (kR_g) R_g \mathrm{d} R_g, \\
& = C_0 R_{g0}^2 J_0 (kR_{g0}).
\end{split}
\end{equation}
Inserting equation (\ref{a}) into equation (\ref{c}) yields an integral over two Bessel functions, which can be shown to give a final expression for the time evolution of the distribution of particles in guiding centre space that follows

\begin{equation}
C(R_g, t) \propto \exp \left[{-\frac{R_g'^2}{4Dt}} \right],
\end{equation}
where $R_g' = R_g(t) - R_g(0)$, and $D=\sigma ^2 / 2 t$, where $\sigma$ is the dispersion in the guiding centre radius distribution. To evaluate the diffusion coefficient, $D$, we define a time window given by an initial and final time:  $\Delta T = T_f - T_i$, and calculate the quantity $R_g'$ for each star particle in the disc. From the assumption that the star particles are initially located in a narrow range of angular momentum, the dispersion is calculated from the final guiding centre distribution only, giving the diffusion coefficient as:

\begin{equation}
D \approx \frac{\overline{R_g'^2}}{2 \Delta T}.
\label{eqdc}
\end{equation} 
It is important to note that the description of radial migration as a diffusion process is valid only for time-scales less than the diffusion time-scale, as discussed by \citet{BCP11}. We find a suitable time window to be of order a dynamical time ($\sim200$ Myr), which is consistent with the adopted time-scale measured by \citet{BCP11}. 

For each simulation, we choose four instances in time at which to centre a time window: $7$, $5$, $3$ and $1$ Gyr look back time. These times are chosen such that they are late enough in the galaxy evolution stages such that there is a well-formed stellar disc with bar/spiral structure. At the beginning of each time window, we calculate the guiding centre of each disc star particle\footnote{We calculate the guiding centre by first computing the radial profile of the $z$-component of angular momentum for circular orbits, $|\mathbf{L}_{z,c}| = |\mathbf{R}\times \mathbf{V}_c|$. The guiding centre is then given by the radius at which the magnitude of the $z$- component of angular momentum of a given star particle intersects this curve.} and divide the disc star particles in guiding centre space into annuli of 0.5 kpc width. At the end of each time window, we calculate the guiding centre distribution of the same star particles that were found in a given radial annulus at the beginning of the window, and apply equation (\ref{eqdc}) to the distribution in order to calculate a diffusion coefficient in each annulus. In this way, we obtain a radial profile of diffusion coefficients. 

\begin{figure*}
\centering
\includegraphics[scale=2.,trim={1cm 0.5cm 0 0.5cm},clip]{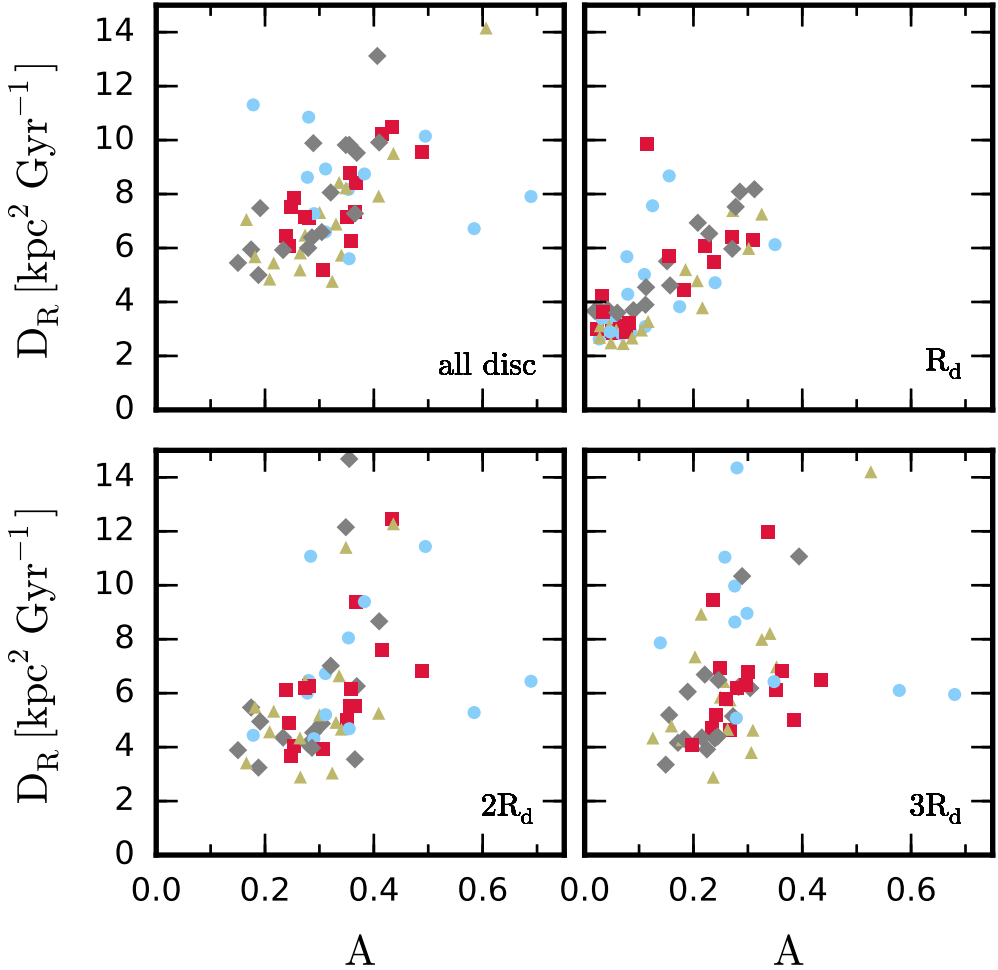}
\caption{{\bff Diffusion coefficients calculated from equation (\ref{eqdc}) as a function of non-axisymmetric structure amplitude, calculated at $t_{\rm lookback}=7$ (circles), 5 (squares), 3 (triangles) and 1 Gyr (diamonds). \emph{Top left}: Coefficients radially averaged from the halo centre to $R=15$ kpc, plotted as a function of the quadrature sum of the spiral and bar amplitudes. The remaining panels show the diffusion coefficients at the radius indicated, plotted against the bar amplitude (top right), bar and spiral quadrature sum (bottom left) and spiral amplitude (bottom right), in accordance with the contributing amplitudes at that radial position.} }
\label{AD}
\end{figure*}

At each of the four short time windows defined above, we characterise the strength of radial migration over the disc as the radial average of the diffusion coefficients, which we plot against the combined strength of bar and spiral structure, defined to be equal to $(A_{\mathrm{b}}^2+A_{\mathrm{sp}}^2)^{1/2}$. This is done for all simulations and shown in {\bff the top-left panel of} Fig.~\ref{AD}. It is clear that there is a positive correlation between the strength of non-axisymmetric structure and the {\bff disc-averaged} amount of migration that takes place in a disc. {\bff To explore the radial dependence of the diffusion coefficients, Fig.~\ref{AD} shows the diffusion coefficients at $1$, $2$, and $3$ disc scale lengths for the same time windows, plotted against the bar, spiral$+$bar and spiral amplitudes, respectively. A particularly tight relationship at $1R_{\rm d}$ is evident, which highlights bar-driven migration in the inner disc region. The correlation is still present at $2$ and $3R_{\rm d}$, though it appears steeper with diffusion coefficients that reach up to $15$ kpc$^2$ Gyr$^{-1}$ in some cases. The increased migration at larger radii may be an indication that satellite induced migration occurs at some times, which is likely more frequent at earlier, more active times of the evolutionary history, as hinted by the lack of correlation for the earliest time window shown at $3R_{\rm d}$. However, the general good correlations shown in all panels of Fig.~\ref{AD} } indicates that the radial migration from spiral arms and a bar indeed occurs in the simulations.  

\begin{figure*}
\centering 
\includegraphics[scale=1.2,trim={0 0 0.6cm 0},clip]{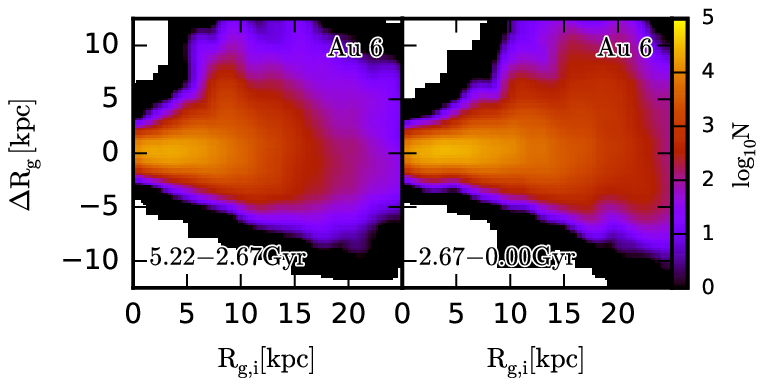} \includegraphics[scale=1.2,trim={1.25cm 0 0 0},clip]{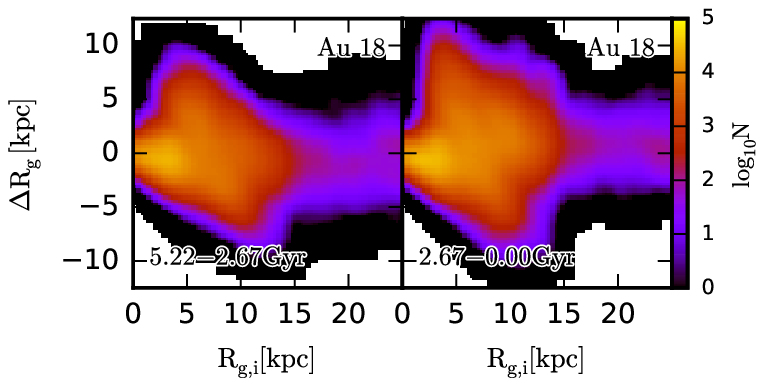} \\ 
\includegraphics[scale=1.2,trim={0 0 0.6cm 0},clip]{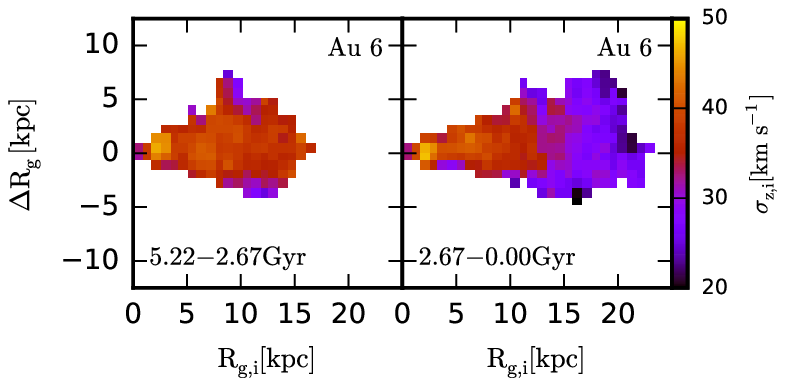}  \includegraphics[scale=1.2,trim={1.25cm 0 0 0},clip]{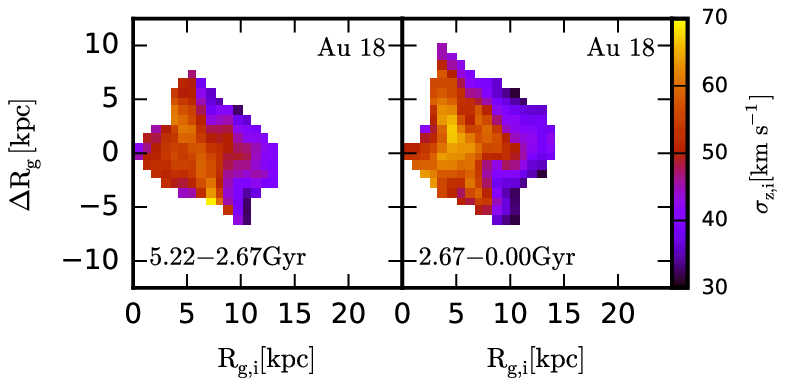} \\  
\includegraphics[scale=1.2,trim={0 0 0.6cm 0},clip]{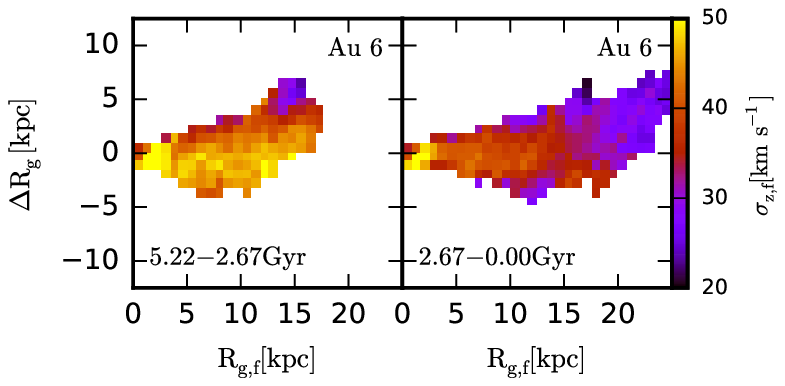}  \includegraphics[scale=1.2,trim={1.25cm 0 0 0},clip]{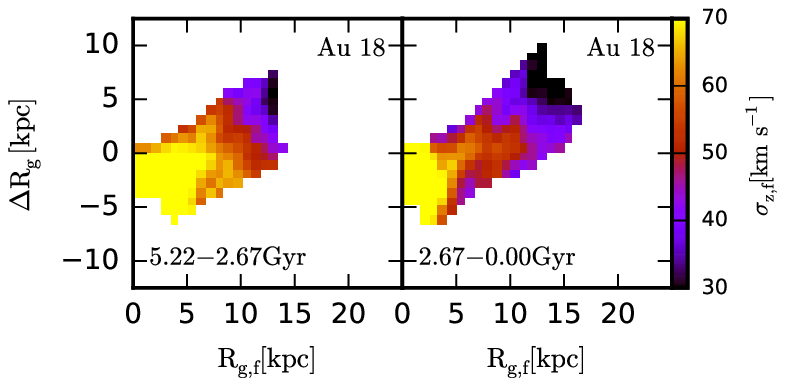}
\caption{The evolution of disc star particle guiding centres at various time periods for the spiral only galaxy in {\bff Au 6} (left two columns) and the barred-spiral galaxy in Au 18 (right two columns). \emph{Top row}: 2D number density histogram of disc star particles in the $\Delta R_g$ vs. $R_{g,i}$ plane, where $R_{g,i}$ is the guiding centre at the beginning of the time window, and $\Delta R_g$ is the change in guiding centre radius between the beginning and end of the time window. The time windows are $\sim 2.6$ Gyr in size, and together span the evolution from $z=0.5$ to present day. \emph{Middle row}: As the top row, but colour-coded according to the vertical velocity dispersion in each bin at the beginning of the time window. \emph{Bottom row}: As the middle row, but colour coded according to the vertical velocity dispersion at the end of the time window, and the $x$-axis now plots the final guiding centre.}
\label{h1618}
\end{figure*}

\subsection{Effect of radial migration on vertical disc structure}

In this section, we examine the effects of radial migration on the vertical structure of the disc, which we study on both short and cosmological time-scales. On the short time-scales, we track the evolution of disc star particles present at the beginning of a given time window only, which enables a clear comparison between star particles that radially migrate with those that do not at a given epoch. On the cosmological time-scale, we consider all star particles present at $z=0$, as a series of coeval populations, which allows us to probe the effects of radial migration on the final disc distribution throughout the formation history of the galaxy.

\subsubsection{Short-term radial migration}
\label{strm}

We divide the period of time from $t_{\rm lookback} = 5.2$ Gyr to present day ($z=0.5$) to ($z=0$) into two equal size time windows, and compute the change in guiding centre radius of each disc star particle between the beginning and end of each time window as a function of the initial guiding centre. The first two panels of Fig.~\ref{h1618} show the smoothed 2D histograms of the disc star particles in this plane for {\bff Au 6} as an example of a spiral-only disc. The histograms illustrate that radial migration occurs over a radial range of between $\sim5$ and $\sim 15$ kpc in the first time window, and expands to up to $\sim 20$ kpc for the second time window as a result of the disc growth that takes place. {\bff These radial ranges are consistent with that of the spiral structure, though note the feature at $R_g > 15$ kpc with $\Delta R_g > 5$ kpc, that may be partly driven by satellite interaction.}

The first two panels of the second row of Fig.~\ref{h1618} show the star particles plotted on the same plane as the top row, but binned into grid cells of side length $0.8$ kpc and colour-coded according to the initial vertical velocity dispersion. These plots indicate, particularly at later times, that the star particles that radially migrate the most are kinematically cool in comparison to those star particles that migrate less/do not migrate. The reason for the preferential migration of kinematically cool stars is that they undergo smaller vertical excursions from the disc plane compared to the kinematically hot stars, and therefore a larger fraction of their orbital period is spent in the disc where the bar and spiral structure are hosted. This result is consistent with simulations of isolated galaxy discs set up in idealised conditions \citep[][]{VC14}. The first two panels of the bottom row of Fig.~\ref{h1618} shows the distribution of the same star particles as the second row binned according to $\Delta R_g$ and final guiding centre radius, $R_{gf}$, colour-coded to the vertical velocity dispersion at the end of the time window. This plot indicates that star particles that gained angular momentum remain cool with respect to the star particles that do not migrate, i.e., those with $\Delta R_g \sim 0$, and contrarily, star particles that lost angular momentum increase their intrinsic vertical velocity dispersion, which in the earlier time window is boosted above that of the local, non-migrating star particles.

The right two columns of Fig.~\ref{h1618} show the same set of plots as the left two columns, but for the barred-spiral disc of Au 18. A typical signature of the bar is clearly identified as a diagonal ridge centred on a guiding centre of about $\sim 7$ kpc, which is caused by the radial migration that occurs close to the bar co-rotation radius: material in the proximity of this resonance that lies inside and outside is torqued outwards and inwards, respectively. The second row shows that the bar is able to cause star particles with high vertical velocity dispersions to migrate, though the same trend seen for Au 25 is still discernible. However, the bottom panels show that stars that migrated outwards undergo significant intrinsic kinematic cooling with respect to their kinematic state at the beginning of the time window \citep[see also][]{GKC11}. This behaviour is expected from the (particle ensemble average) conservation of vertical action \citep[e.g.,][]{SoS12,Min12}, which predicts that vertical energy decreases proportional to $\mathrm{exp} [ -R / 2R_s]$ \citep[e.g.,][]{Min12,Rok12}. In the case of Au 18, the outward migrated star particles cool to such a degree as to be kinematically cooler than non-migrator star particles at their final guiding centre radius {\bff (see also Fig.~\ref{monsigz})}, despite their high original velocity dispersion. 

\begin{figure*}
\includegraphics[scale=1.5]{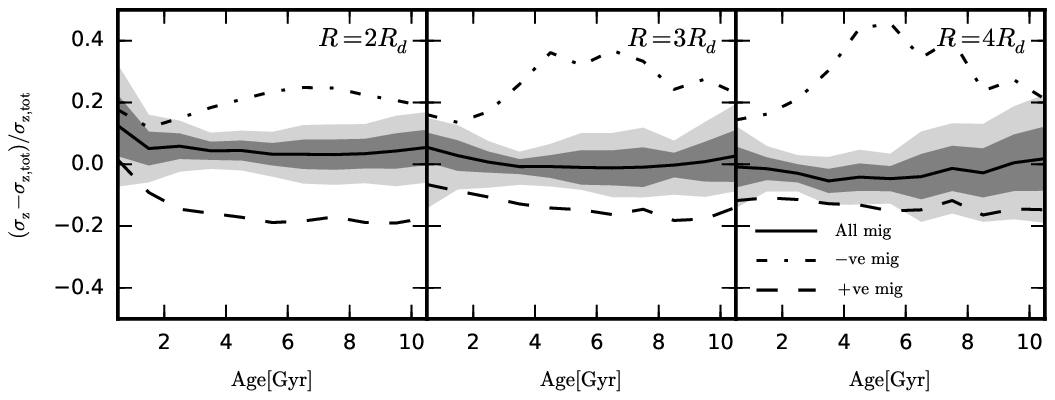}\\
\includegraphics[scale=1.5]{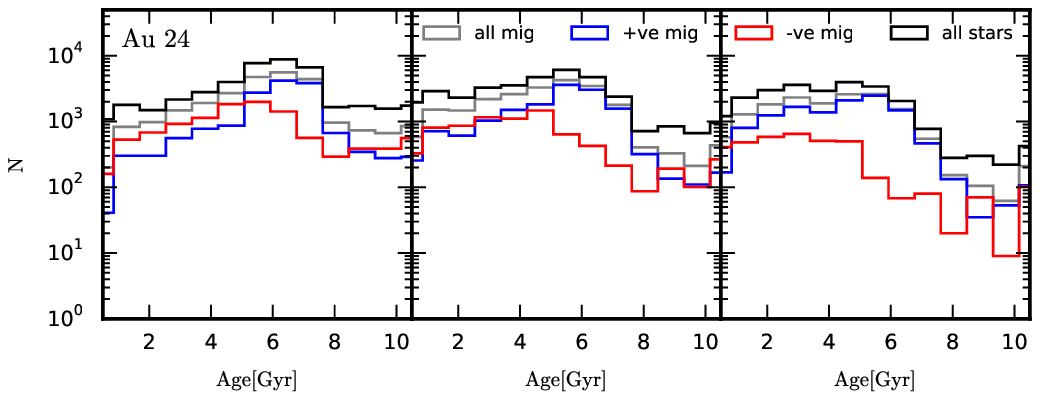}
\caption{\emph{Top row}: Fractional contribution of migrated star particles to the vertical velocity dispersion found at a given radius (indicated in the panel) as a function of stellar age, for each simulation. The mean values of the simulation suite for all migrated star particles is given by the solid black line, and the 1 and 2-$\sigma$ regions are indicated by the dark and light grey shaded regions, respectively. The mean values of the simulation suite for the negative and positive migrated star particles considered separately are given by the dot-dashed and dashed black lines, respectively. \emph{Bottom row}: Histograms of the age distribution of all star particles (black), all migrated (grey), outward migrated (blue) and inward migrated (red) star particles found at the radius indicated in the top row for Au 24.}
\label{trsigz}
\end{figure*}

\begin{figure}
\includegraphics[scale=1.15]{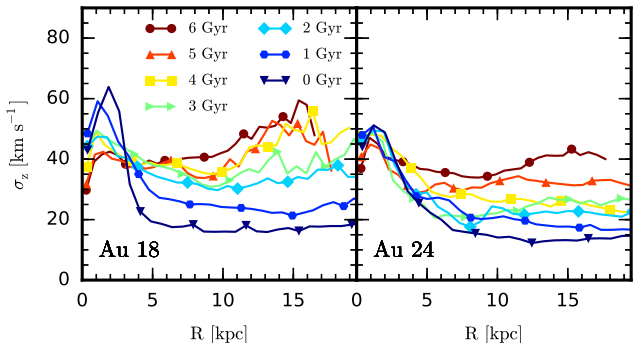}
\caption{Radial profiles of the vertical velocity dispersion for star particles less than 1 Gyr old at a series of lookback times, for Au 18 (left) and Au 24 (right).}
\label{yng}
\end{figure}

\subsubsection{Radial migration on cosmological time-scales}

In this section we quantify the extent to which radial migration on cosmological time-scales influences the present day disc structure. We divide all {\bff in-situ} star particles at present day into age bins of width $1$ Gyr, and calculate their present day guiding centre values, $R_g(t=0)$, and their birth radii, $R_{\rm birth}$, which we assume to be nearly circular (and therefore approximately equal to the guiding centre). We define $\Delta R_g = R_g(t=0) - R_{\rm birth}$. For each coeval population at a given present day radius, we define star particles that have migrated outward (inward) as those that were born inside (outside) of the radius and that satisfy $\Delta R_g > 1$ kpc ($\Delta R_g < -1$ kpc).  At a set of three present day radii: $2$, $3$ and $4$ times the disc scale length, the present day vertical velocity dispersion of the migrated star particles in each coeval bin is calculated, and normalised as a fraction of the total present day vertical velocity dispersion in each coeval bin, $\sigma _{z,\rm tot}$, at each given radius.

The top row of Fig.~\ref{trsigz} shows the results for the innermost (left panel), intermediate (middle panel) and outermost (right) radius for all simulations. The simulation sample mean of the fraction of vertical velocity dispersion of all migrated star particles relative to all the star particles in a given age bin is shown by the solid black curve, and the 1- and 2-$\sigma$ regions are indicated by the dark and light shaded regions, respectively. The mean fraction of vertical velocity dispersion of star particles that migrated inwards only (negative migrated star particles) and star particles that migrated outwards only (positive migrated star particles) are shown by the dot-dashed and dashed curves, respectively. Clearly, a general trend is that negative (positive) migrated star particles have a heating (cooling) effect on the disc at all radii, which appear to cancel out at most ages and produce a sample mean effect of migration close to zero. Note that the scatter among the simulations is very small across most ages, which indicates that this trend is statistically robust. At the innermost radius, the sample mean for overall migration indicates that radially migrated star particles contribute to an increase of less than $\sim 10 \%$ in the local vertical velocity dispersion, whereas no effect is observed in the two outermost radii. The cause of this trend can be seen in the bottom row of Fig.~\ref{trsigz}, which shows histograms of the star particle age distribution at each radius for different groups of star particles for {\bff Au 24}. It is clear that inward migrated star particles make up {\bff a significant fraction} of the migrated star particles at the innermost radius, {the kinematic heating effect of which} dominates over the cooling of the outward migrated star particles. {\bff At larger radii, the positive migrated star particles become more important,} which produces very little {\bff overall} effect on the vertical velocity dispersion \citep[see also][]{Min12,MCM14}. This appears to be a general trend among the simulation suite.

Although the mean effect of radial migration is near zero for most of the stellar ages, migrated young star particles (Age $<2$ Gyr) exhibit an enhanced velocity dispersion relative to the local star particles, which is most pronounced at the  innermost radius: of order $\sim15\%$ on average with a relatively large scatter among the halos. While this feature is caused primarily by the dominance of inward migrated star particles (which is emphasised at young ages) at the innermost radius, it is accompanied by an up-turn in the trend for young positive migrated star particles. This indicates that at later times, positive migrated star particles originate from a kinematically hotter inner disc relative to the outer disc, the contrast of which is highest at $2 R_s$ (compared to 3 and 4$R_s$). The origin of this contrast may be attributed to a comparatively quiescent period of evolution at late times with much less disturbance from minor mergers and satellite interactions (which are thought to be more effective in the outer disc regions), or that the orbits of newborn star particles become kinematically cooler with time, particularly in the outer disc regions (or a combination thereof). To investigate the latter possibility, we plot in Fig.~\ref{yng} the radial profile of the vertical velocity dispersion of disc star particles younger than $1$ Gyr at a series of times from $t_{\rm lookback}=6$ Gyr to present day, for Au 18 and {\bff Au 24} (which are typical examples). It is clear that stars born at later times are born on progressively colder orbits, {\bff particularly for star particles in the outer disc regions.} This leads to a radial profile that drops steeply between radii of 0 and 5 kpc, and is flat outside 5 kpc. The relatively high dispersion in the inner regions means that star particles found at $R=2R_s$ that originate from radii less than $\sim 5$ kpc are kinematically hotter than local star particles, despite the kinematic preference for low velocity dispersion star particles to migrate (see Fig.~\ref{h1618}). However, star particles younger than $\sim 2$ Gyr constitute a small fraction of the total migrated star population, particularly at the innermost radii, which indicates that these star particles do not have a significant effect on the overall vertical kinematic structure. Note that the age distribution peaks at $\sim 7$ Gyr, $\sim 6$ Gyr and $\sim 5$ Gyr at the innermost, intermediate and outermost radius, respectively, which is a clear indication of inside-out formation.

The result that radial migration has little impact on disc heating/thickening is consistent with several other studies that have used simulations to investigate radial migration \citep{Min12,MCM14,VC14,MMF14,VC15}. Furthermore, although each mono-abundance population exhibits a degree of flaring which is larger in older populations \citep{MMS15}, the lack of any contribution from radial migration to disc heating in the outer regions of the disc indicates that the flaring is not caused by star particles migrating from the inner disc regions, {\bff as is found to be the case for discs in isolation \citep[e.g.,][]{Min12}.} This seems to be related to the respective cooling and heating effects of positive and negative migrated star particles in our simulations, whose behaviour is opposite to what is found by \citet{Min12,MMF14}. Our results are more consistent with those of \citet{MCM14,MCM14b}, who showed that the discs of galaxies that experience mergers are perturbed more in the outer regions than the inner regions, such that positive migrator star particles that travel to the outer regions are relatively cooler than the local perturbed stars, and vice versa for negative migrators. In this case, radial migration acts to suppress disc flaring.

\section{Other sources of vertical heating}
\label{secos}

We have shown above that radial migration cannot account for any disc heating seen in the stellar discs of our simulation suite, and that the most plausible source of internal disc heating is the bar, whose strength correlates well with the disc-averaged heating of a coeval population in some simulations. In this section we investigate two possible sources of disc heating that do not relate to non-axisymmetric structure: adiabatic heating and external perturbations from sub-halos/satellite galaxies.

\begin{figure}
\includegraphics[scale=2.]{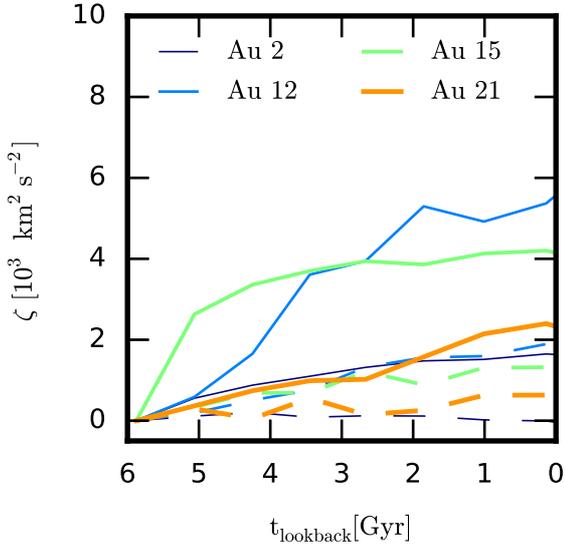}
\caption{The vertical velocity dispersion of a coeval population of star particles born at $7 \mathrm{Gyr} > t_{\rm lookback} > 6 \mathrm{Gyr}$ as a function of time (solid lines). The contribution to the velocity dispersion inferred from an increase in the mid-plane density and assuming a constant scale height is represented by the dashed lines. {\bff Au 2} is representative of the other halos not shown.}
\label{adia}
\end{figure}

\subsection{Adiabatic heating}

\begin{figure*}\hspace{0.0mm}
\includegraphics[scale=1.15,trim={0 2.cm 0 0},clip]{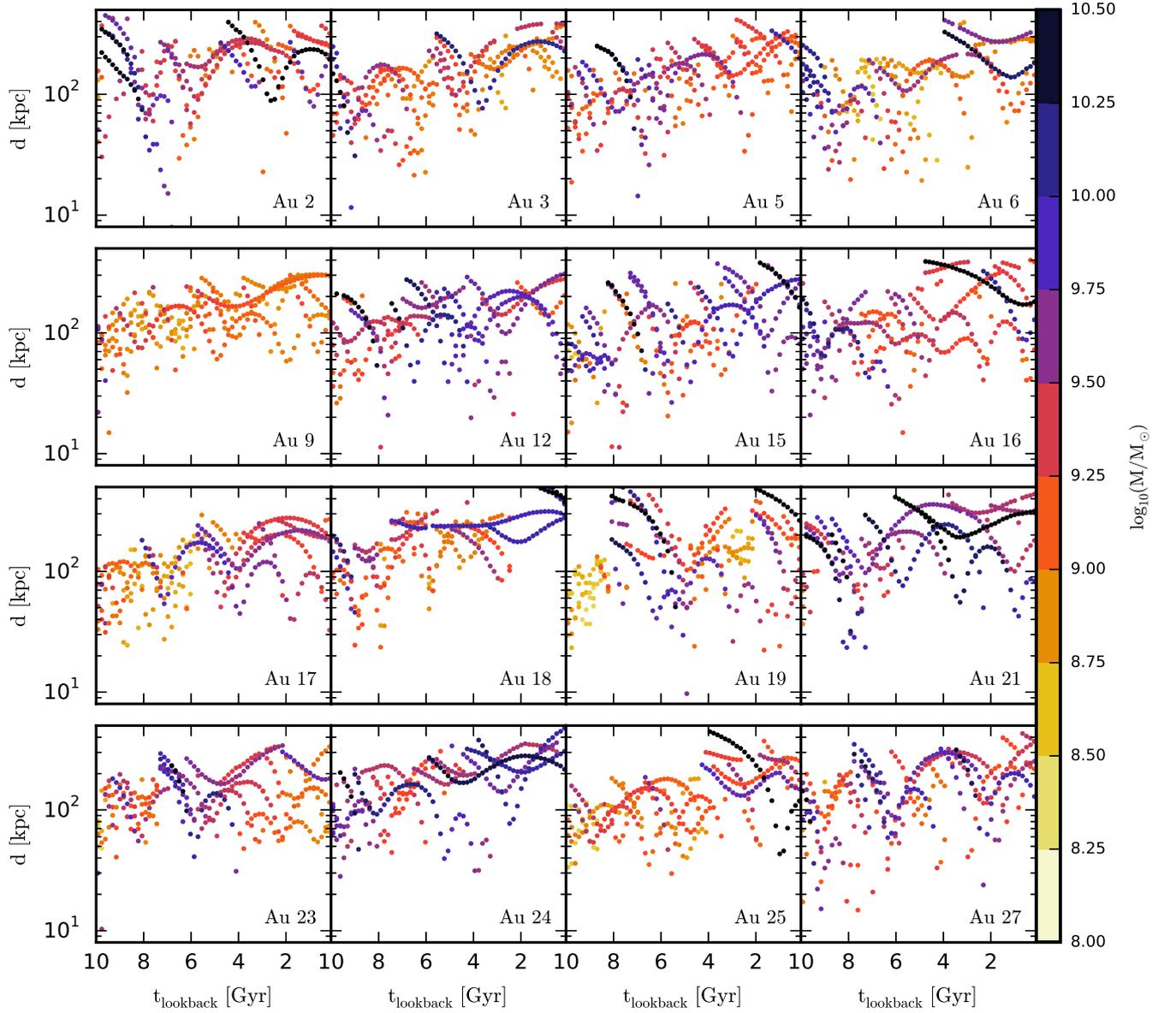}
\caption{The distance from the halo centre of the five most massive sub-halos as a function of time, for each halo. The colours indicate the total mass of each sub-halo.}
\label{satdist}
\end{figure*}

As discussed in section \ref{sims}, the disc in many of the simulations exhibits signs of late-time growth. Gradual increases in the mid-plane surface density of baryonic material slowly increases the gravitational potential of stars that oscillate above and below the disc plane, and therefore injects vertical energy into pre-existing stars adiabatically \citep{J92,VKH10}.  

In Fig.~\ref{adia} we show the evolution of the vertical velocity dispersion of star particles born between $7 > t_{\rm lookback} > 6$ Gyr found within a radial range of $1R_s$ and $3R_s$, normalised to the vertical velocity dispersion at $t_0 = 6$ Gyr, for a sub-sample of simulations. To determine how much of the heating of this coeval population is accounted for by mid-plane disc growth, we calculate the surface density of the baryonic material (stars and gas) in the same radial range, within $3$ kpc vertical distance of the mid-plane. We make the assumption that the velocity dispersion scales with the square root of the surface density, $\sigma _z \propto \sqrt{h_z \Sigma}$, which is true, for example, in the case of an isothermal density distribution \citep{S42,B84}. We adopt the conservative assumption that the scale height of a population of star particles remains constant with time subject to an increase in the mid-plane density \citep[see also][]{MMF14}, which places and upper limit on resulting velocity dispersion increases. The evolution of the normalised vertical velocity dispersion {\bff squared} inferred from mid-plane growth is then given by 

\begin{equation}
\Big[ \sigma ^2 _z (t) - \sigma ^2 _z (t_0) \Big]_{ad} = 2 \pi G h_z (t_0) \Big( \Sigma (t) - \Sigma (t_0) \Big),
\end{equation}
which is shown by the dashed curves in Fig.~\ref{adia}. For each simulation shown the increase in vertical velocity experienced by the coeval population is not fully explained by adiabatic heating. {\bff The largest fraction of the absolute kinematic heating that can be accounted for by adiabatic heating is found in Au 15 to be $\sim 40\%$, although the absolute heating is low.} The other simulations in our suite show a contribution comparable to {\bff Au 2}, which indicates that adiabatic heating from disc mid-plane growth is not a major source of disc heating.

\begin{figure*}\hspace{0.0mm}
\centering
\includegraphics[scale=1.3,trim={1.cm 0 0 0},clip]{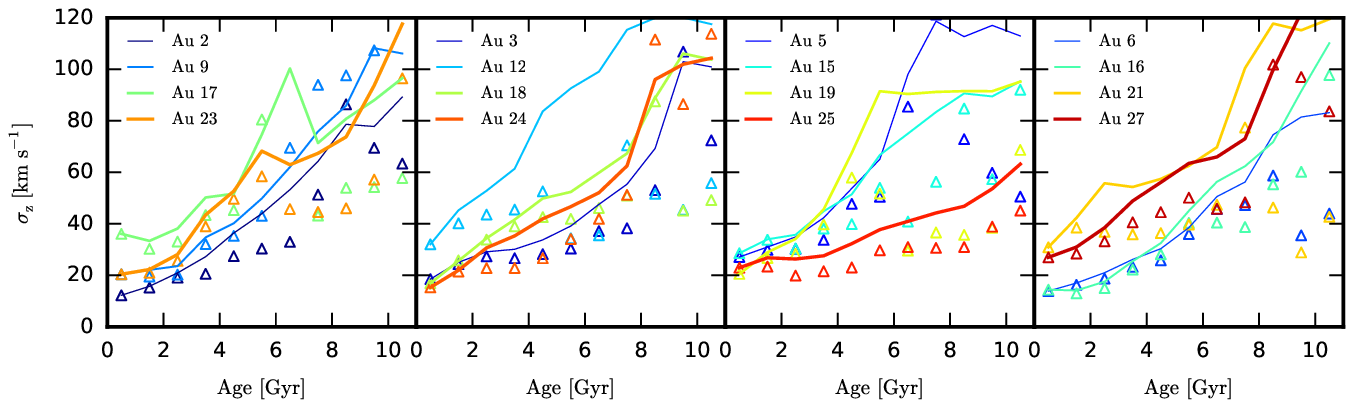}
\caption{The vertical velocity dispersion of star particles, in the radial range that spans $2$-$3R_s$, of coeval star particles at present day (solid curves) and at birth (triangles), for all simulations. For illustration purposes, we spread the plots across four panels.}
\label{avr}
\end{figure*}

\subsection{External perturbations}
\label{extp}

Perturbations from orbiting sub-halos/satellite galaxies have been shown to be significant to the vertical structure of the main galaxy \citep{PB10,GMV12,GMO13,WBC14,GWM15,MJ15,DMV15}. In Fig.~\ref{satdist}, we plot for each simulation, the galacto-centric distance of the five most massive sub-halos at a given time from a redshift of $t_{\rm lookback}=10$ Gyr to present day. The points are colour-coded according to sub-halo total mass. {\bff Significant interactions at early times are indicated in halos Au 2 and Au 21 (both at $t_{\rm lookback} \sim 8$ Gyr), which are consistent with the sudden extreme kinematic heating at these times in both the inner and outer disc regions, as highlighted in Fig~\ref{sbheat}.} It is clear that Au 19 experiences an encounter with a satellite of total mass greater than $\sim 10^{10.25}$ $\rm M_{\odot}$ at $t_{\rm lookback}\sim 5$ Gyr, which subsequently in-falls and experiences significant mass loss between $t_{\rm lookback} \sim 5$ and $3$ Gyr. This timing is consistent with the extreme heating that occurs in this halo shown in Fig.~\ref{sbheat}. A similar interaction occurs in Au 12, which experiences dramatic periodic heating events that seem to be associated with multiple perturbations from satellites of masses $\sim 10^{10}$ $\rm M_{\odot}$ over the period between $5 > t_{\rm lookback} > 2$ Gyr {\bff (in addition to a bar)}. {\bff Other halos that experience similar multiple interactions after $t_{\rm lookback} = 6$ Gyr include} Au 15 and Au 21, which do not host strong bars and undergo appreciable heating, appear to be similarly harassed by satellites. In contrast, discs that host a very weak/no bar and experience very quiescent evolution from $z=1$ to present day, such as those in Au 6 and Au 25, exhibit near constant vertical velocity dispersion with time (Fig.~\ref{sbheat}), consistent with the idea that a bar and satellite interactions are primary sources of disc heating. This indicates also that satellites with masses less than $\sim 10^{9.5} \rm M_{\odot}$ are negligible in the context of disc heating.

\section{The Age-Velocity dispersion relation (AVR)}
\label{secav}

The AVR is a crucial observable relation that contains information about the nature of dynamical heating processes that take place over the evolutionary history of a galaxy. For example, secular heating processes such as those discussed in this paper are expected to yield a smooth AVR, whereby disc stars are heated gradually over cosmological time periods. In contrast, more violent mechanisms such as mergers are likely to be imprinted into the AVR as sharp jumps in velocity dispersion over narrow age ranges, indicative of rapid heating. In this section, we study the AVRs in our simulation suite and correlate their shapes to the dynamical histories of the galaxies, in order to assess the impact of secular heating mechanisms on the overall disc structure.  

In Fig.~\ref{avr} we show the AVR, calculated from star particles in a radial range that spans $2$-$3R_s$, for each simulation. As expected, we see a variety of AVR shapes among the simulation sample. The impact of sub-halos/satellites is clearly seen as sharp jumps in the AVR, which is particularly clear at ages between 4 and 6 Gyr in Au 19, which saturates for star particles older than 6 Gyr (which constitutes star particles that are present before the interaction). In contrast, for many halos the AVR is smooth up to ages of 10 Gyr, consistent with the expectation that internal secular evolution gradually heats the disc on cosmological time-scales. 

However, it is interesting to note that even those discs that do not host a strong bar or experience prominent satellite interactions, for example Au 5, 6 and 16, can still exhibit a steep AVR despite the apparent lack of disc heating mechanisms. This suggests that the heating mechanisms we have studied so far do not exclusively shape the AVR. To investigate the extent to which dynamical heating of pre-existing star particles shapes the AVR, we compare the velocity dispersion of each coeval population at present day with that at their birth time (shown by the triangles in Fig.~\ref{avr}). In many cases, this is also an increasing function of age, i.e., newborn star particles are born on orbits of decreasing velocity dispersion as the galaxy evolves. In particular, we see that the AVR of the quiescent halos mentioned above appears to be entirely caused by newborn star particles that become progressively colder with time. Furthermore, this leads to the overall vertical velocity dispersion and scale height of the disc to decrease with time. The only discs that do not become kinematically cooler with time are those of Au 12 and 19, which experience the strongest satellite interactions. This indicates that internal secular heating mechanisms are no match for upside-down formation of the disc, and that strong satellite interactions have a comparable but opposite effect. These results highlight that upside-down disc formation \citep{BK13,SBR13,BRS15} occurs on some level ubiquitously, which has important implications for thick disc formation.

\section{Resolution Study}
\label{secrs}

In any $N$-body simulation, there is some level of artificial heating caused by two-body interactions of point-mass particles, in which particles are scattered onto orbits of increased random energy. Studies such as \citet{Fu11} and \citet{Se13} have shown that the amount of numerical heating depends strongly on the resolution of the simulation: numerical heating is higher in simulations of lower particle number. It is generally accepted that numerical heating in simulations in which the disc is modelled with $N>10^6$ particles is suppressed to an acceptable level, whereas those of lower particle number are in danger of being dominated by it. Although the simulations presented in this paper are above this threshold ($\sim 1-3$ million disc particles), it is important to ensure that resolution-dependent numerical heating does not play an important role in the evolution of the system. 

\begin{figure}
\includegraphics[scale=2.]{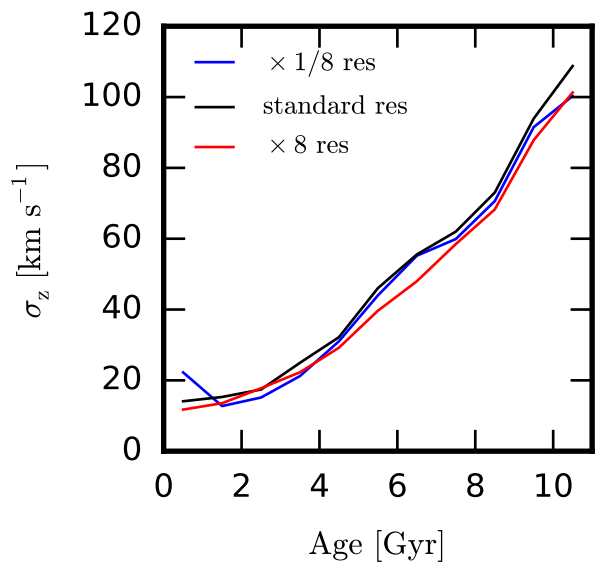}
\caption{The age-velocity dispersion relation for Au 16 at the standard resolution (black), 8 times higher resolution (red) and 8 times lower resolution (blue).}
\label{res}
\end{figure}

To this end, we perform two further simulations of Au 16: one high resolution run in which the particle number is increased by a factor of $8$ ($\sim 10$ million disc star particles), and a low resolution run in which the particle number is decreased by a factor of $8$ ($\sim 10^5$ disc star particles). We compare the disc heating in these simulations over the last 10 Gyr of evolution from inspection of the age-velocity dispersion profile shown in Fig.~\ref{res}. Overall, the AVR profiles are very similar, with deviations in vertical velocity dispersion less than $8.0$ km s$^{-1}$ for any age. This indicates that the heating effects in our simulations are not dominated by numerical heating effects, but instead by physical heating mechanisms. 

\section{Conclusions}
\label{seccon}

In this paper, we performed a suite of cosmological-zoom simulations with the state of the art $N$-body magneto-hydrodynamics code \textlcsc{AREPO}, which models many aspects of galaxy formation and is successful in reproducing a broad range of observed properties of late-type galaxies \citep{MPS14,PMS14} and of the global galaxy populations \citep{VGS14,VGS14b,GVS14}. The simulation suite comprises 16 galaxies, which are shown to exhibit a variety of well-resolved bar and spiral structure in well-defined stellar discs, some of which continue to grow until $z=0$. We follow the evolution from $z=1$ to present day, and first characterise the spiral and bar structure evolution. We then examine the impact of secular processes linked to non-axisymmetric structure and external perturbations on the disc heating of pre-existing stars, and assess their relative importance to the overall evolution of the vertical disc structure. We come to the following conclusions:

\begin{itemize}
\item{} We characterised the strength of bar and spiral structure with the amplitudes of Fourier modes calculated from the density distribution. We compared the time evolution of their strength with the global vertical energy evolution of a coeval population of star particles and found evidence for a positive correlation with bar strength. In some cases, the presence of a bar appears to be the sole driver of vertical heating. On the contrary, spiral arms do not appear to cause much disc heating, however this may be because of a lack of larger mass disc particles (such as GMCs) to efficiently divert planar motions to the vertical direction. 

\item{} We characterised the strength of radial migration as a set of radially dependent diffusion coefficients that describe the broadening of guiding centre ($z$-component of angular momentum) distributions of star particles on dynamical time-scales. These diffusion coefficients reveal a positive correlation between the strength of radial migration and that of non-axisymmetric structure at several epochs in time, which confirms that radial migration from bars and spiral arms occurs in the simulations.

\item{} We analyse the effects of radial migration on the vertical velocity dispersion profile on both short ($\sim$1 Gyr) and cosmological time-scales. The short time-scale analysis reveals that star particles that take part in radial migration in a given time window are kinematically cooler than star particles that migrate less/do not migrate. This confirms, for the first time, that the recent findings of isolated disc simulations \citep[e.g.,][]{Min12,GKC11,VC14} hold in a cosmological context. 

\item{} The overall effect of radial migration on the present day vertical disc is sub-dominant for all coeval populations, which suggests that radial migration cannot thicken pre-existing stellar distributions \citep[contrary to what is reported by, e.g.,][]{SB09b,Rok12}. {\bff It follows that} radial migration {\bff does} not have a flaring effect on the vertical {\bff disc structure (scale heights and density profiles to be presented in Grand et al. in prep)}. This is contrary to what is found in some previous studies \citep[e.g.,][]{Rok12}, and may be explained by the relatively flat radial profile of the vertical velocity dispersion in our simulations, which appears to reverse the kinematic effect of positive/negative migrated star particles on the local particles. 

\item{} Many of the discs exhibit late time growth in the mid-plane surface density, which leads to some degree of adiabatic heating in the vertical direction. However, we find that in most cases, this does not have a significant effect on the vertical velocity dispersion. There are only three halos in which this effect produces a vertical heating more than a few percent over the period from $z=0.5$ to $0$. 

\item{} A significant source of disc heating in at least a quarter of the simulations appears to be perturbations associated with sub-halos/satellites of masses $>10^{10} \rm M_{\odot}$. The most significantly heated discs in our sample, Au 12 and Au 19, show bursts of heating at times associated with pericentre passages of satellites. In the latter case, a large fraction of the total disc heating appears to be caused by such an interaction. 

\item{} In nearly all of the simulations, star particles are born on orbits that become kinematically cooler with time, especially after $z=0.5$ (and in some cases from $z > 1$). We find that this effect is dominant over the secular heating mechanisms, which leads to an overall vertical disc that cools and becomes thinner with time. This evolution is in favour of the recently proposed upside-down disc formation \citep{HBG11,BK13,SBR13,BRS15}. There are only two discs that are not dominated by this mechanism: those that experience the strongest satellite interactions. 

\item{} Finally, we performed a resolution test that compares our standard resolution with both a higher and lower resolution simulation, and find that the AVR does not change significantly between them. This indicates that the heating mechanisms in the simulations do not have a numerical origin.

\end{itemize}
 
Overall, our results highlight that the most significant heating episodes arise from sub-halo interactions in the form of fly-bys and minor mergers, though in our simulations sample they are, {\bff at least after $z \sim 1$,} less prevalent than the secular heating provided by a bar. An arguably more important issue is related to how the kinematic properties of newborn star particles evolve with time, an effect that appears to cool the disc and dominate the shape of the AVR to a large extent. The origin(s) of this behaviour should be tied to the evolution of the state of star-forming gas, which may cool with time owing to lower stellar feedback at later times. The evolution of the star-forming gas may also depend upon the interstellar medium physics and how it is modelled, and therefore it is possible that stellar kinematics could be used as a diagnostic for both ISM and viable feedback models.

The conclusions of this paper are relevant to thick disc formation from stars that were born in-situ. We have not followed the evolution of star particles accreted from orbiting sub-halos, which may impact also the stellar distribution and contribute to a thick disc. Furthermore, we have not attempted to explicitly identify a thick disc and track its formation history. Such an investigation should include detailed chemical analysis of the stellar distribution in combination with kinematics and structural parameters. We reserve this investigation for a future study.

\section*{acknowledgements}
RG acknowledges very helpful and constructive comments from and discussion with Daisuke Kawata, and useful comments from Ivan Minchev, Carlos Vera-Ciro and Marie Martig. RG and VS acknowledge support by the DFG Research Centre SFB 881 `The Milky Way System' (subproject A1). DJRC acknowledges STFC studentship ST/K501979/1. This work has also been supported by the European Research Council under ERC-StG grant EXAGAL- 308037. Part of the simulations of this paper used the SuperMUC system at the Leibniz Computing Centre, Garching, under the project PR85JE of the Gauss Centre for Supercomputing. This work used the DiRAC Data Centric system at Durham University, operated by the Institute for Computational Cosmology on behalf of the STFC DiRAC HPC Facility `www.dirac.ac.uk'. This equipment was funded by BIS National E-infrastructure capital grant ST/K00042X/1, STFC capital grant ST/H008519/1, and STFC DiRAC Operations grant ST/K003267/1 and Durham University. DiRAC is part of the National E-Infrastructure.

\bibliographystyle{mnras}
\bibliography{GG1d4R1.bbl}

\end{document}